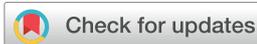

Check for updates

# Structural investigation of doubly-dehydrogenated pyrene cations†

Sanjana Panchagnula, [ID] ★[ab] Jordy Bouwman, [ID][a] Daniël B. Rap, [ID][c] Pablo Castellanos,[ab] Alessandra Candian, [ID][bd] Cameron Mackie,[b] Shreyak Banhatti,[e] Sandra Brünken, [ID][c] Harold Linnartz [ID][a] and Alexander G. G. M. Tielens [ID][b]

The vibrationally resolved spectra of the pyrene cation and doubly-dehydrogenated pyrene cation ($C_{16}H_{10}^{\bullet+}$; Py$^+$ and $C_{16}H_8^{\bullet+}$; ddPy$^+$) are presented. Infrared predissociation spectroscopy is employed to measure the vibrational spectrum of both species using a cryogenically cooled 22-pole ion trap. The spectrum of Py$^+$ allows a detailed comparison with harmonic and anharmonic density functional theory (DFT) calculated normal mode frequencies. The spectrum of ddPy$^+$ is dominated by absorption features from two isomers (**4,5-ddPy$^+$** and **1,2-ddPy$^+$**) with, at most, minor contributions from other isomers. These findings can be extended to explore the release of hydrogen from interstellar PAH species. Our results suggest that this process favours the loss of adjacent hydrogen atoms.

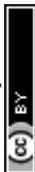

## 1 Introduction

The aromatic infrared bands (AIBs) – a series of broad emission features observed in the near- to mid-infrared spectra of many astronomical sources in interstellar and circumstellar regions – are widely accepted to be caused by polycyclic aromatic hydrocarbons (PAHs).[1,2] Free-floating PAH molecules are excited following the absorption of (vacuum) ultraviolet (UV) photons, triggering complex intramolecular vibrational energy redistribution that results in the emission of infrared (IR) radiation. The IR features are most prominent in areas with strong UV radiation fields such as H II regions, reflection and planetary nebulae, and in the diffuse interstellar medium (ISM).[2,3]

In interstellar regions illuminated by strong UV fluxes, interstellar PAHs tend to be photoionised.[4,5] In addition, the strong UV flux may lead to additional fragmentation of the formed PAH cations through loss of H, $H_2$, or small hydrocarbons.[5,6] Loss of hydrogen from PAHs results in derivatives such as dehydrogenated PAH cations.[7,8] Astrochemical models show that the hydrogenation state of a PAH depends on the local interstellar UV radiation field (leading to hydrogen-loss), and the atomic hydrogen density (which drives re-hydrogenation of the dehydrogenated radical).[4,7,9] In UV-rich regions or regions of low atomic hydrogen density such as clouds near massive stars, PAHs are predicted to exhibit a high degree of dehydrogenation.[4,7] The extent of hydrogen loss also depends strongly on the size of the molecule, with smaller PAHs being subject to more dehydrogenation.[5,9] While hydrogen-loss is the dominant fragment channel, small PAHs ($<16$ carbon atoms) may exhibit a (minor) carbon-loss channel. However, for pericondensed PAHs, including pyrene, loss of $C_2H_2$ is inhibited because a key transition state is too high in energy.[10] Indeed, experiments reveal that for PAHs in the astrophysically-relevant size range (50–100 carbon atoms), fragmentation is dominated by hydrogen-loss until most – if not all – hydrogen atoms are lost.[8,10,11] For completeness, we mention that in regions with high atomic hydrogen density and low UV fields, interstellar PAHs can be superhydrogenated (i.e. some or all of the edge carbon atoms have acquired an additional hydrogen atom).[7,9] Here, we focus on dehydrogenated PAHs as they are particularly relevant in regions of massive star formation.[7]

Precision measurements of the AIBs (strongest at 3.3, 6.2, 7.7, 8.6 and 11.2 µm) reveal subtle changes in band profiles and peak positions.[12,13] These changes must have a molecular origin and reflect that PAHs in space can exist in several forms[14,15] with varying charge states,[16,17] possibly as fragments of larger precursors[18] or with hetero-atom substitutions in the carbon skeleton.[19] Dehydrogenated PAHs in particular have gained interest from the astronomical community as they may be involved as by-products in the formation of molecular hydrogen in regions illuminated by strong UV fields,[20,21] and in the first step towards the conversion of PAHs to cages and fullerenes in space.[11] The exact mechanism by which hydrogen is formed from aromatic molecules has not yet been experimentally determined, and the structures of the resulting dehydrogenated PAHs are also unknown.

[a] Laboratory for Astrophysics, Leiden Observatory, Leiden University, PO Box 9513, 2300 RA Leiden, The Netherlands. E-mail: panchagnula@strw.leidenuniv.nl

[b] Leiden Observatory, Leiden University, PO Box 9513, 2300 RA Leiden, The Netherlands

[c] Radboud University, Institute for Molecules and Materials, FELIX Laboratory, Toernooiveld 7, NL-6525ED Nijmegen, The Netherlands

[d] Van't Hoff Institute for Molecular Sciences, University of Amsterdam, PO Box 94157, 1090 GD Amsterdam, The Netherlands

[e] I. Physikalisches Institut, Universität zu Köln, Zülpicher Str. 77, 50937 Köln, Germany

† Electronic supplementary information (ESI) available. See DOI: 10.1039/d0cp02272a

Experimental challenges such as low molecular densities in molecule-specific detection methods have made it difficult to measure and assign the mid-IR spectra of large PAH cations and radicals in the gas phase, particularly when investigating charged fragmentation products. Action spectroscopy has been widely applied to overcome these challenges.[6,22] Infrared multi-photon dissociation (IRMPD) spectroscopy using strong infrared light sources such as free electron lasers in combination with ion traps offers an effective spectroscopic tool capable of recording spectra using very low molecular abundances; ion traps are typically filled with only a few thousand ions.[23] However, IRMPD also comes with a number of disadvantages, most notably a non-linear response that shifts band positions and alters relative band intensities.[24,25]

Recent advancements in cryogenic techniques are one means of addressing these issues. Infrared predissociation (IRPD) spectroscopy uses the rare gas tagging messenger technique in combination with tunable infrared radiation to provide spectroscopic information on the cold ion-rare gas complex following a single-photon absorption process. For IRPD, observed intensities more closely resemble calculated absorption cross-sections than in the case of IRMPD.[26] This predissociation technique, using IR and NIR photons, has recently been applied to successfully study the spectra of a large number of PAH, PAH photofragment, and fullerene cations.[27–30]

Pyrene (Fig. 1) is a highly stable and commercially available species representing the smallest pericondensed PAH. It has therefore been the subject of many experimental studies as a model system for PAHs when studying their physical chemistry. While small compared to the astrophysically-relevant size range, pyrene exhibits several of the characteristic hydrocarbon edge sites present in interstellar PAHs (Fig. 1) and as such it has become a prototypical species in astronomical model studies of PAHs. The first measurement of IR emission from a gaseous PAH cation (generated by electron impact ionisation) was achieved with pyrene; the spectrum showed relative intensities consistent with astrophysical observations of the AIBs, as well as unidentified features that may be attributed to dehydrogenated pyrene species.[31] The IR absorption features of the pyrene cation have been recorded in matrix isolation experiments and assigned on the basis of theory work.[32] These have triggered IRMPD work[23] onto which the present work extends.

In this work we present the IRPD spectra of the pyrene cation ($Py^+$; $C_{16}H_{10}^{\bullet+}$) and its doubly-dehydrogenated cation ($ddPy^+$; $C_{16}H_8^{\bullet+}$) recorded using the neon-tagging messenger technique. We explore the dehydrogenation process resulting in $ddPy^+$ and, with the aid of theoretical computations, propose the most likely mixture of isomeric candidates that constitute the $ddPy^+$ sample.

## 2 Methods

### 2.1 Experimental

The infrared spectra of neon-tagged pyrene ($Ne\cdots C_{16}H_{10}^{\bullet+}$, $m/z = 222$) and doubly-dehydrogenated pyrene ($Ne\cdots C_{16}H_8^{\bullet+}$, $m/z = 220$) were recorded using the cryogenic 22-pole ion trap instrument FELion at the Free Electron Laser for Infrared Experiments Laboratory (FELIX, Radboud University).[33] A detailed description of the FELion end-station is provided elsewhere,[26] and here only a brief description of the apparatus and information specific to the target ions is given. One major advantage of this method, compared to the previously applied IRMPD, is that intensity ratios of recorded spectra do not suffer from a highly non-linear multi-photon response and allow for better comparisons with theoretical predictions.

A solid pyrene sample (Sigma-Aldrich, 99%) was heated to reach a vapour pressure of $1–2 \times 10^{-5}$ mbar and dissociatively ionised by electron impact ionisation at electron energies of 20–40 eV in an ion source. A 100 ms long pulse of the resulting cations ($Py^+$ or $ddPy^+$) was extracted from the source and filtered for the mass of interest by a first quadrupole mass selector (mass spectrum given in Fig. S1, ESI†). The ions were guided into the 22-pole ion trap which was maintained at a fixed temperature of 6.5 K by means of a closed-cycle helium refrigerator. They were then cooled close to the ambient temperature by a 140 ms long $1:3$ Ne:He gas mixture pulse at a number density of $\sim 10^{15}$ cm$^{-3}$ provided by a pulsed piezo valve. The buffer gas inlet was triggered 20 ms before the ions were admitted into the trap, and lasted for 20 ms after the ion pulse. Under these conditions, around 10% of the primary ions ($Py^+$ or $ddPy^+$) formed weakly bound complexes with neon.

The so-formed complexes were typically stored in the trap for 1.6 s, and during this time irradiated by several laser pulses of the defocused IR radiation provided by the free electron laser FEL-2 operated in the 5.6–18 μm (550–1800 cm$^{-1}$) range at a repetition rate of 10 Hz. FEL-2 delivered up to 30 mJ per macropulse into the 22-pole trap, and the FEL was optimised for narrow bandwidths of full-width-at-half-maximum (FWHM) of 0.5–1% of the used wavelength.

An IRPD spectrum was recorded by mass-selecting and counting the target PAH–neon complex ions and varying the laser wavelength. The absorption of a resonant IR photon caused the complex to dissociate and led to a depletion in the number of complex ions compared to the baseline number observed off-resonance. To account for saturation, varying baseline, laser pulse energy, and pulse numbers, the signal was normalised prior to averaging, yielding the intensity in units of relative cross section per photon. Three iterations were averaged for every data point and the spectra were binned to a resolution of 1 cm$^{-1}$.

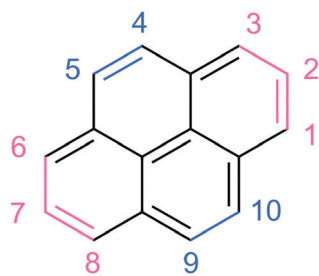

**Fig. 1** Chemical structure of pyrene. The hydrogen-numbering system shown here is employed when labelling isomers of the doubly-dehydrogenated pyrene cation. Blue shows duo hydrogen sites, and pink shows trio hydrogen sites.

## 2.2 Theoretical

The density functional theory (DFT) quartic force field of Py$^+$ was obtained with the Gaussian16[34] software package. The B3LYP[35,36] hybrid functional was used in conjunction with the polarized double-$\zeta$ basis set N07D,[37] which has been established to work well for open-shell species.[38] The resulting calculated force field was supplied as an input for SPECTRO 8 to determine the anharmonic spectrum of the molecule.[18]

Vibrational harmonic spectra of all possible ddPy$^+$ isomers were predicted by DFT computations. Geometry optimisations were performed at the B3LYP/6-311++G(2d,p) level of theory and the calculated harmonic frequencies were uniformly scaled by 0.983 to account for anharmonicity.[39] This empirically-determined scaling factor is in-line with the suggested scaling factor of 0.982 for PAH IR spectra computed with the chosen basis set at the B3LYP level of theory.[40] All modes were convolved using a Gaussian profile with FWHM of 8 cm$^{-1}$ to facilitate comparison with the experimental spectra.

The B3LYP/6-311++G(2,dp) level of theory was chosen as it is particularly well-suited to vibrational frequency studies. For consistency, we also explored the potential energy surfaces connecting the ddPy$^+$ isomers using the same level of theory. As this functional is not optimised for the study of reaction energies and barriers, we have repeated these calculations using the more accurate range-separated functional wB97XD with the 6-311++G(2d,p) basis set for several key isomers and transition states.

## 3 Results and discussion

### 3.1 Pyrene cation, Py$^+$

The black trace in Fig. 2 shows the IRPD spectrum recorded for neon-tagged Py$^+$. For comparison, the theoretical anharmonic and harmonic spectra convolved with a Gaussian profile at a FWHM of 8 cm$^{-1}$ are plotted in blue and pink, respectively (Fig. 2). The calculated band positions are represented by sticks. The experimental spectrum obtained in this work represents a clear improvement on the gas-phase IRMPD spectrum of Py$^+$ available from literature,[23] showing better signal-to-noise ratio, narrower linewidths, and well-resolved features. Peak positions and band assignments are listed in Table 1 along with unpublished gas-phase IRMPD data recorded by our group (experimental details[58] and data are provided in Fig. S2 and S3, ESI†), and matrix data available from the literature.[32] The band positions derived here are consistent with those from the IRMPD study, and improved FWHM values as low as 4 cm$^{-1}$ are realised for isolated absorptions.

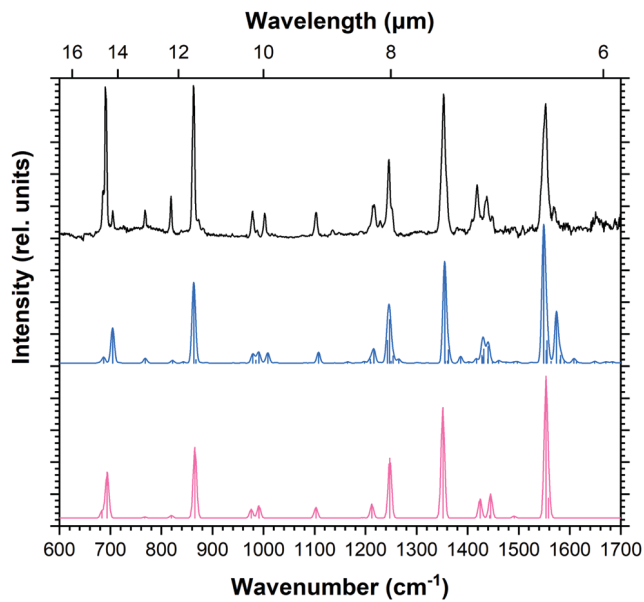

**Fig. 2** IRPD spectrum of Py$^+$ (black trace) with the calculated anharmonic spectrum (blue), and the calculated harmonic spectrum scaled by 0.983 (pink).

IRMPD is a non-linear process in which several photons are required to reach the dissociation threshold. The IRPD technique is a linear process where only one photon is needed to dissociate the complex, since the binding energy of the neon atom is typically only a few hundred wavenumbers. This allows weaker bands to be detected. Most notably, the C–H in-plane bending modes at 978 and 1103 cm$^{-1}$ are prominent in the IRPD spectrum whereas in the IRMPD spectrum only the former could be observed as a weak band. Similarly for the C–H out-of-plane bending modes; while only the bands at 690 and 861 cm$^{-1}$ were observed in the IRMPD data, additional bands were recorded in the IRPD spectrum. The band positions appearing in our IRPD and previous argon matrix data show excellent agreement (deviations ≤4 cm$^{-1}$).[32]

DFT computations for Py$^+$ resulted in a $^2B_{1g}$ electronic ground state with a $D_{2h}$ structure. Band positions resulting from (unscaled) harmonic and anharmonic DFT calculations are listed in Table 1. All fundamental and combination modes ($\nu_i + \nu_j$) with relative intensities higher than 3% are reported. The spectroscopic assignment of all IRPD bands to calculated anharmonic frequencies is straightforward. The agreement between the anharmonic frequencies and the experimental data is excellent despite a minor blue-shift of 2–10 cm$^{-1}$ in the calculations. While the two calculated spectra appear remarkably similar after scaling the harmonic spectra by an empirically-determined factor of 0.983, the anharmonic calculations reveal weak bands (that are absent in the harmonic calculations) which contribute to most features such as the 1408 cm$^{-1}$ band due to the combination of $\nu_{53}$ (704 cm$^{-1}$) and $\nu_{54}$ (687 cm$^{-1}$). The convolved line shapes for the harmonic and anharmonic spectra are similar enough that one can draw an equally convincing match to the observed spectrum from either calculation.

**Table 1** Experimental and theoretical (DFT) vibrational transitions of Py$^+$ [a]

| IRPD (this work) | | | IRMPD[b] | | Argon matrix[c] | | Emission[d] | | Harm. DFT[e] | | Anh. DFT | |
| --- | --- | --- | --- | --- | --- | --- | --- | --- | --- | --- | --- | --- |
| $\nu_{vib}$ | FWHM | $I_{rel}$ | $\nu_{vib}$ | $I_{rel}$ | $\nu_{vib}$ | $I_{rel}$ | $\nu_{vib}$ | $I_{rel}$ | $\nu_{vib}$ | $I_{abs}$ | $\nu_{vib}$ | $I_{abs}$ |
| 686 | 8.7 | 0.58 | | | | | | | 694 | 7.8 | 687 | 7.1 |
| 691 | 3.4 | 1.00 | 680 | 0.15 | 690 | 0.23 | | | 691 | 43.3 | 704 | 43.3 |
| 705 | 4.6 | 0.11 | | | | | | | | | | |
| 768 | 3.9 | 0.13 | | | | | | | 777 | 6.0 | 768 | 5.7 |
| 819 | 3.5 | 0.24 | | | | | | | | | 822 | 3.3 |
| | | | | | | | | | | | 843 | 1.5 |
| 864 | 5.1 | 0.95 | 854 | 0.56 | 861 | 0.27 | | | 871 | 102.8 | 863 | 97.9 |
| 872 | 10.0 | 0.11 | | | 868 | 0.04 | | | | | 868 | 4.1 |
| | | | | | 954 | 0.07 | | | | | | |
| 978 | 6.0 | 0.95 | 989 | 0.13 | 977 | 0.10 | | | 993 | 11.1 | 979 | 10.8 |
| 987 | 4.3 | 0.24 | | | | | | | | | 985 | 3.4 |
| | | | | | | | | | | | 991 | 13.0 |
| 1002 | 5.2 | 0.15 | | | | | | | 1001 | 14.6 | 1008 | 12.3 |
| 1103 | 5.7 | 0.16 | 1095 | 0.05 | 1102 | 0.09 | | | 1123 | 13.5 | 1108 | 13.3 |
| 1136 | 7.5 | 0.15 | | | | | | | | | 1164 | 1.2 |
| | | | | | | | | | | | 1198 | 2.0 |
| | | | 1212 | 0.63 | 1189 | 0.04 | 1189 | 0.01 | | | 1209 | 4.6 |
| 1210 | 10.4 | 0.05 | | | | | | | | | | |
| 1216 | 7.3 | 0.18 | | | 1216 | 0.13 | 1209 | 0.10 | 1237 | 17.4 | 1216 | 16.9 |
| 1229 | 10.9 | 0.07 | 1235 | 1.00 | | | | | | | 1229 | 6.9 |
| 1244 | 12.7 | 0.83 | | | 1245 | 0.38 | 1243 | 0.18 | | | 1243 | 27.8 |
| 1246 | 8.8 | 0.44 | | | | | | | | | 1247 | 53.4 |
| | | | | | | | | | | | 1250 | 2.1 |
| 1252 | | 0.51 | | | 1255 | 0.05 | | | 1265 | 75.9 | 1254 | 8.8 |
| | | | | | 1345 | 1.25 | | | | | 1265 | 5.0 |
| 1353 | 10.2 | 0.86 | 1378 | 0.49 | 1357 | 1.12 | 1343 | 1.20 | 1385 | 138.3 | 1355 | 121.0 |
| | | | | | | | | | | | 1356 | 2.1 |
| | | | | | | | | | | | 1361 | 2.6 |
| | | | | | 1362 | 0.18 | 1366 | 0.22 | | | 1362 | 16.4 |
| | | | | | | | | | | | 1386 | 8.1 |
| 1408 | 5.8 | 0.07 | 1405 | 0.65 | | | 1403 | 0.53 | | | 1402 | 1.2 |
| 1419 | 8.9 | 0.29 | | | 1421 | 0.18 | 1418 | 0.46 | | | 1418 | 5.0 |
| 1426 | | 0.21 | | | | | | | | | 1428 | 9.3 |
| 1437 | 12.0 | 0.22 | | | 1440 | 0.11 | | | | | 1429 | 28.6 |
| | | | | | | | | | | | 1438 | 5.0 |
| 1449 | 4.5 | 0.10 | | | | | | | 1467 | 27.0 | 1440 | 20.4 |
| | | | | | | | | | | | 1461 | 3.0 |
| | | | | | | | | | | | 1475 | 1.2 |
| | | | | | | | | | | | 1490 | 1.4 |
| | | | | | | | | | | | 1497 | 1.5 |
| 1551 | 9.7 | 0.82 | 1538 | 1.00 | 1553 | 1.00 | 1553 | 1.00 | 1586 | 181.4 | 1549 | 159.1 |
| | | | | | | | | | | | 1554 | 15.7 |
| | | | | | | | | | | | 1555 | 4.8 |
| | | | | | | | | | | | 1555 | 27.4 |
| | | | | | | | | | | | 1564 | 2.0 |
| 1570 | 9.7 | 0.14 | | | | | | | 1591 | 28.6 | 1574 | 62.6 |
| | | | | | | | | | | | 1581 | 4.0 |
| | | | | | | | | | | | 1582 | 9.5 |
| | | | | | | | | | | | 1589 | 1.7 |
| | | | | | | | | | | | 1608 | 5.1 |
| | | | | | | | | | | | 1613 | 1.3 |
| | | | | | | | | | | | 1649 | 1.9 |
| | | | | | | | | | | | 1671 | 1.2 |
| | | | | | | | | | | | 1684 | 1.4 |

[a] Frequencies are given in cm$^{-1}$ and calculated intensities ($I_{abs}$) are given in km mol$^{-1}$. In the case of anharmonic calculations, only modes with relative intensities higher than 3% are reported. [b] See Fig. S2 and S3, ESI. [c] Hudgins *et al.*[32] [d] Kim *et al.*[31] [e] No frequency scaling is used.

The recorded vibrational spectra presented in this study show no unaccounted bands, leading us to the conclusion that no other electronically-excited states are present. The precise effect of neon-coordination on the IRPD spectrum of Py$^+$ is unknown but is expected to be small (a few cm$^{-1}$).[41] Theoretical calculations of the various neon tag coordinations are, given the good match between the DFT calculated spectrum of the untagged pyrene cation and the experimental data, out of the scope of this work.

### 3.2 Doubly-dehydrogenated pyrene cation, ddPy$^+$

IRPD spectroscopy was also employed to record the mid-IR spectrum of the doubly-dehydrogenated pyrene radical cation (Fig. 3). There are a greater number of features, several of which appear broader than those recorded for Py⁺. The doubly-dehydrogenated form of pyrene has many more active modes than the fully hydrogenated pyrene cation due to the loss of symmetry. This leads to apparently broader structures that in reality consist of multiple overlapping bands. The spectrum also shows greater variation between the recorded and predicted intensities of the bands when compared to the pyrene molecule, as will be discussed later. This is likely a result of saturation of the features which causes weaker modes to appear more intense. The spectrum is fitted using multiple Gaussian components with typical FWHM values of 10 cm⁻¹, giving a total of 39 bands listed in Table 3.

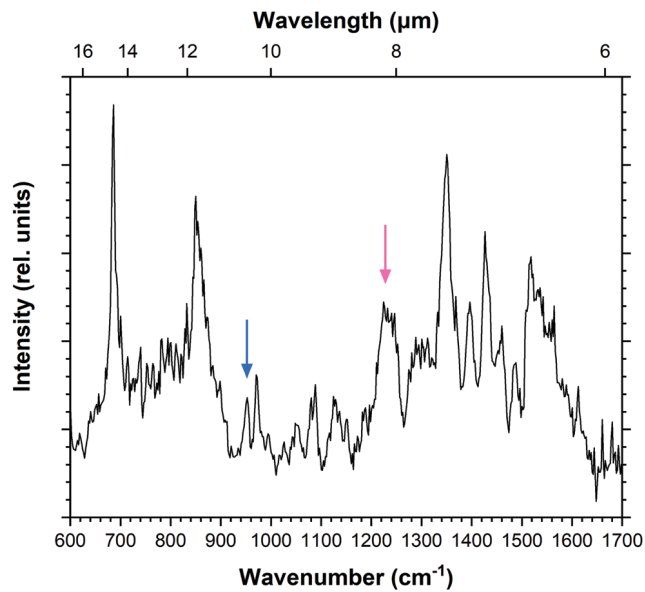

**Fig. 3** IRPD spectrum of ddPy⁺. Arrows highlight bands at 952 cm⁻¹ (blue) and 1245 cm⁻¹ (pink) which are relevant for saturation–depletion measurements.

**Table 2** DFT calculated electronic ground states (EGS), symmetries, and energies (ΔE, given in eV) of all possible ddPy⁺ isomers[a]

| Isomer | Config. ($\pi^{14}\cdots$) | EGS | Symm. | $\Delta E^{b}$ |
|---|---|---|---|---|
| **4,5-ddPy⁺** | $\pi^1\sigma^2$ | $^2B_1$ | $C_{2v}$ | 0.00 |
| **1,2-ddPy⁺** | $\pi^1\sigma^2$ | $^2A''$ | $C_s$ | 0.05 |
| **1,3-ddPy⁺** | $\pi^1\sigma^2$ | $^2A_2$ | $C_{2v}$ | 0.28 |
| 1,5-ddPy⁺ | $\pi^1\sigma^1\sigma^1$ | $^2A''$ | $C_s$ | 1.08 |
| 1,8-ddPy⁺ | $\pi^1\sigma^1\sigma^1$ | $^2A''$ | $C_s$ | 1.11 |
| 2,7-ddPy⁺ | $\pi^1\sigma^1\sigma^1$ | $^2A_u$ | $D_{2h}$ | 1.11 |
| 2,4-ddPy⁺ | $\pi^1\sigma^1\sigma^1$ | $^2A''$ | $C_s$ | 1.14 |
| 2,5-ddPy⁺ | $\pi^1\sigma^1\sigma^1$ | $^2A''$ | $C_s$ | 1.14 |
| 1,4-ddPy⁺ | $\pi^1\sigma^1\sigma^1$ | $^2A''$ | $C_s$ | 1.15 |
| 1,7-ddPy⁺ | $\pi^1\sigma^1\sigma^1$ | $^2A''$ | $C_s$ | 1.16 |
| 4,9-ddPy⁺ | $\pi^1\sigma^1\sigma^1$ | $^2A_u$ | $C_{2h}$ | 1.16 |
| 4,10-ddPy⁺ | $\pi^1\sigma^1\sigma^1$ | $^2B_1$ | $C_{2v}$ | 1.17 |
| 1,9-ddPy⁺ | $\pi^1\sigma^1\sigma^1$ | $^2A''$ | $C_s$ | 1.19 |
| 1,10-ddPy⁺ | $\pi^1\sigma^1\sigma^1$ | $^2A''$ | $C_s$ | 1.22 |

[a] Isomers are sorted in decreasing order of stability with hydrogen-loss sites numbered according to Fig. 1. [b] Energy difference with respect to the most stable isomer, **4,5-ddPy⁺**.

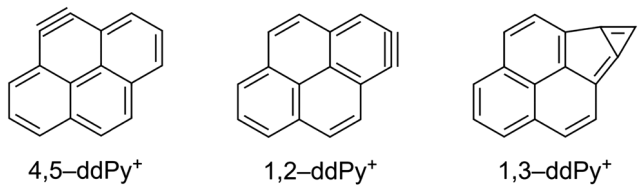

**Fig. 4** Chemical structures of **4,5-ddPy⁺**, **1,2-ddPy⁺**, and **1,3-ddPy⁺**.

In order to assign the experimental vibrational spectrum of the ddPy⁺ species, DFT structure optimisations and harmonic frequency calculations were performed on fourteen possible isomers in the doublet state using the B3LYP/6-311++G(2d,p) level of theory. The relative energies of the fourteen ddPy⁺ isomers are listed in Table 2. The good agreement in peak frequency between the shifted harmonic DFT calculations for Py⁺ gives confidence that scaled harmonic calculations for ddPy⁺ will suffice. While all isomers with $\pi^{14}\cdots\pi^1\sigma^1\sigma^1$ configuration have approximately the same relative energies (0.14 eV range), the most stable isomers are the three with a $\pi^{14}\cdots\pi^1\sigma^2$ configuration: **4,5-ddPy⁺**, **1,2-ddPy⁺**, and **1,3-ddPy⁺** (shown in Fig. 4). The numbers refer to the positions where the H-atoms are missing (see Fig. 1). Further analysis using the range-separated wB97XD/6-311++G(2d,p) level of theory reveals that the relative energies of these species are essentially identical (see Table S0, ESI†).

For these three isomers, two hydrogen atoms are removed from adjacent or nearby carbon atoms. In the case of hydrogen removal from adjacent carbon atoms (**4,5-ddPy⁺** and **1,2-ddPy⁺**), the dehydrogenated carbons become sp-hybridised and an additional π-bond is formed, creating a triple bond as the lowest energy structure in a doublet state. With **1,3-ddPy⁺**, all carbon atoms remain sp²-hybridised and an additional σ-bond is formed between carbons 1 and 3, transforming a 6-membered carbon ring into a cyclopropenyl unit (a bicyclic structure with a 5-membered ring and a 3-membered ring) which has been shown to stabilise cationic molecules.[42] As an aside, we note that this latter species is even better stabilised as a dicationic molecule with charges delocalised separately in a phenalene cation and a cyclopropenyl cation.[42]

Computed vibrational spectra of the six most stable ddPy⁺ isomers plotted together with the experimental ddPy⁺ spectrum are shown in Fig. 5. The calculated vibrational transitions of these isomers are listed in Table 3 along with the experimentally recorded ddPy⁺ absorption bands. A scaling factor of 0.983 that was derived from the Py⁺ spectrum is applied to the computations. Because of the structural similarity of the isomers, they generally exhibit only small changes between the positions and relative intensities of the main transitions.

For all fourteen considered ddPy⁺ isomers, the computed IR spectra show strong absorptions at the characteristic C–H out-of-plane bending modes at ~680 and ~840 cm⁻¹. The C–H in-plane bending modes at ~1200 and ~1300 cm⁻¹ are also visible, as is the strong, broad C=C ring stretch at ~1500 cm⁻¹. **1,2-ddPy⁺** has a strong C–H in-plane bending mode at 1140 cm⁻¹ that the other isomers do not exhibit. Likewise, **1,3-ddPy⁺** has a unique transition at 790 cm⁻¹ originating from an in-plane deformation mode involving the triangular structure formed by the carbon bonds. This is the strongest transition for this isomer. A similar bicyclic structure was determined as the most stable isomer for the $C_6H_4^+$ radical (*m*-benzyne) cation with CASPT2, and DFT, CCSD(T) and MP2 calculations,[43,44] supporting that the cyclopropenyl deformation is real and not simply an artefact of the computational methods used. **1,3-ddPy$^+$** was also investigated with the BP86 and wB97XD functionals using the same 6-311++G(2d,p) basis set to ensure that the 790 cm$^{-1}$ band was real.

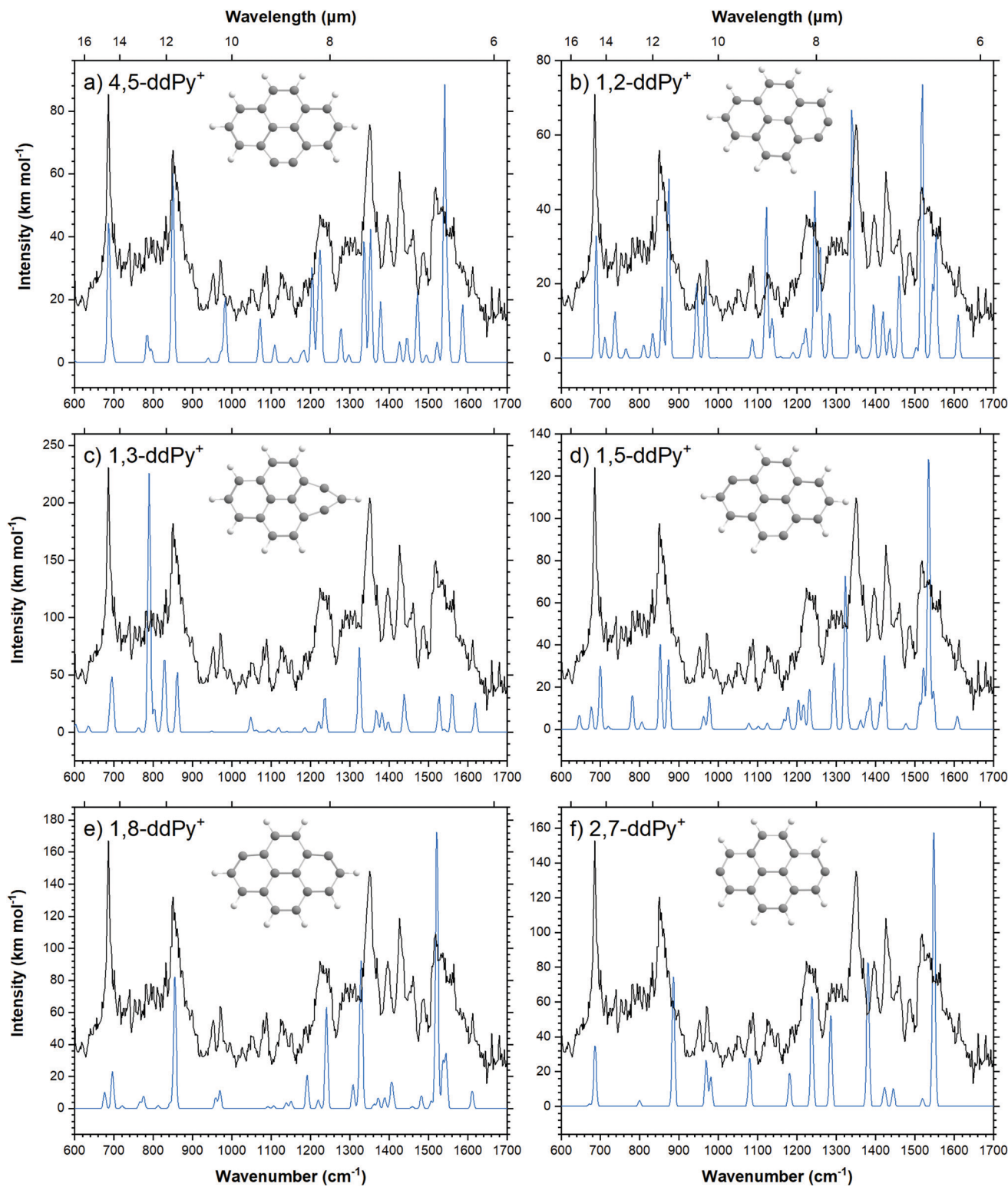

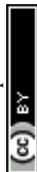

**Fig. 5** Calculated spectra (blue) of the six most stable isomers of ddPy$^+$ (see Table 3) performed with B3LYP/6-311++G(2d,p), scaled by 0.983. The recorded spectrum is shown in black.

**Table 3** Experimental (IRPD) bands of ddPy$^+$ and scaled (0.983) theoretical (DFT) vibrational transitions of the six lowest-energy ddPy$^+$ isomers[a]

| IRPD (this work) | | | DFT calculations | | | | | | | | | | | |
| | | | 4,5-ddPy$^+$ | | 1,2-ddPy$^+$ | | 1,3-ddPy$^+$ | | 1,5-ddPy$^+$ | | 1,8-ddPy$^+$ | | 2,7-ddPy$^+$ | |
| $\nu_{vib}$ | FWHM | $I_{rel}$ | $\nu_{vib}$ | $I_{abs}$ | $\nu_{vib}$ | $I_{abs}$ | $\nu_{vib}$ | $I_{abs}$ | $\nu_{vib}$ | $I_{abs}$ | $\nu_{vib}$ | $I_{abs}$ | $\nu_{vib}$ | $I_{abs}$ |
|---|---|---|---|---|---|---|---|---|---|---|---|---|---|---|
| 580 | 4.5 | 0.10 | | | | | | | | | | | | |
| 586 | 3.1 | 0.11 | | | | | 591 | 5.5 | | | | | | |
| 600 | 5.2 | 0.12 | 599 | 1.0 | 595 | 2.6 | 602 | 7.1 | | | | | | |
| 686 | 8.9 | 1.00 | 686 | 44.3 | 688 | 32.9 | | | | | | | 686 | 34.7 |
| 700 | 9.7 | 0.36 | | | | | 695 | 48.3 | 700 | 23.0 | | | | |
| 714 | 11.6 | 0.23 | | | 711 | 5.7 | | | | | | | | |
| 730 | 16.0 | 0.19 | | | 737 | 12.5 | | | | | | | | |
| 764 | 4.2 | 0.09 | | | 763 | 2.3 | 764 | 3.8 | | | | | | |
| 782 | 5.2 | 0.16 | 786 | 8.6 | | | | | 781 | 16.1 | | | | |
| 794 | 10.4 | 0.17 | 794 | 4.4 | | | | | | | | | | |
| 800 | 3.6 | 0.10 | 797 | 3.5 | | | | | | | | | 800 | 3.4 |
| 810 | 13.9 | 0.22 | | | 808 | 3.4 | | | | | | | | |
| 828 | 9.7 | 0.16 | | | | | 829 | 63.0 | | | | | | |
| 832 | 2.5 | 0.13 | | | 834 | 6.6 | | | | | | | | |
| 850 | 20.3 | 0.88 | 849 | 60.2 | 857 | 19.1 | | | 852 | 40.2 | | | | |
| 874 | 9.7 | 0.33 | | | 874 | 48.2 | | | 873 | 32.8 | | | | |
| 954 | 12.4 | 0.16 | | | 946 | 20.0 | | | | | 958 | 6.8 | | |
| 970 | 9.1 | 0.25 | | | 968 | 19.3 | | | 977 | 15.5 | 970 | 11.1 | 969 | 26.5 |
| 1054 | 17.3 | 0.09 | | | | | 1048 | 13.2 | | | | | | |
| 1080 | 11.4 | 0.17 | 1071 | 14.0 | | | | | | | | | | |
| 1088 | 6.7 | 0.20 | | | 1085 | 5.2 | 1090 | 1.8 | | | | | | |
| 1129 | 9.6 | 0.16 | | | 1122 | 40.6 | | | 1125 | 2.9 | | | | |
| 1150 | 8.5 | 0.13 | 1149 | 1.6 | | | | | | | 1150 | 4.4 | | |
| 1188 | 14.1 | 0.13 | 1184 | 4.0 | | | 1186 | 4.1 | | | 1191 | 21.1 | | |
| 1225 | 23.4 | 0.52 | 1224 | 35.8 | 1222 | 8.0 | 1220 | 9.3 | | | | | | |
| 1252 | 15.7 | 0.48 | | | 1245 | 45.0 | | | | | | | | |
| 1350 | 14.4 | 0.83 | 1352 | 42.5 | | | | | | | | | | |
| 1371 | 12.9 | 0.45 | 1378 | 19.3 | | | | | | | | | | |
| 1396 | 16.6 | 0.45 | | | 1393 | 14.3 | 1395 | 8.2 | | | | | | |
| 1426 | 12.4 | 0.69 | 1427 | 6.7 | 1419 | 12.6 | | | 1422 | 35.1 | | | 1422 | 10.8 |
| 1430 | 5.7 | 0.46 | | | 1436 | 7.89 | | | | | | | | |
| 1460 | 7.1 | 0.38 | | | 1459 | 22.1 | | | | | | | | |
| 1486 | 11.2 | 0.21 | 1472 | 22.5 | 1473 | 22.5 | | | 1482 | 7.8 | | | | |
| 1496 | 1.7 | 0.14 | 1492 | 2.3 | | | | | 1512 | 12.6 | | | | |
| 1516 | 15.2 | 0.64 | 1521 | 6.5 | 1519 | 73.7 | | | 1522 | 29.2 | 1520 | 172.5 | 1519 | 4.5 |
| 1539 | 20.4 | 0.57 | 1541 | 88.5 | 1544 | 19.8 | | | | | 1536 | 28.3 | | |
| 1561 | 14.4 | 0.50 | | | | | 1559 | 33.1 | | | | | | |
| 1588 | 13.9 | 0.25 | 1586 | 18.5 | 1587 | 18.4 | | | | | | | | |
| 1613 | 6.0 | 0.23 | | | 1610 | 11.6 | 1619 | 25.9 | | | 1610 | 10.6 | | |

[a] Frequencies ($\nu_{vib}$ and FWHM) are given in cm$^{-1}$ and calculated intensities ($I_{abs}$) are given in km mol$^{-1}$.

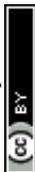

The corresponding band appears at 721 cm$^{-1}$ when scaled by 0.983 for BP86, and 864 cm$^{-1}$ when scaled by an empirical scaling factor of 0.973 (see Fig. S6, ESI†) for wB97XD. This underlines that the calculated band position of the cyclopropenyl vibrational mode is rather sensitive to the choice of computational method (with a shift of 69 cm$^{-1}$ in B3LYP vs. BP86, and 74 cm$^{-1}$ in B3LYP vs. wB97XD), but all methods agree that this mode should have a very strong intensity. It follows that the **1,3-ddPy$^+$** isomer would produce a single, very intense band in the vicinity of 864 cm$^{-1}$, however such a band is conspicuously absent over the relevant frequency range (721–864 cm$^{-1}$) indicated by these different calculations. Therefore, the absence of this distinct band implies that **1,3-ddPy$^+$** may not be an abundant species in our ddPy$^+$ spectrum, which hints at even lower abundances of isomers with higher energy configurations.

On the basis of the theoretical isomeric spectra it is evident that no single isomer is capable of explaining the full measured ddPy$^+$ spectrum, and instead a mixture of isomers is required.

Three possible mixtures of isomers were considered and compared with the experimental spectrum to gain further insight:

(A) A 1 : 1 mixture of **4,5-ddPy$^+$ : 1,2-ddPy$^+$**

(B) A 1 : 1 : 0.5 mixture of **4,5-ddPy$^+$ : 1,2-ddPy$^+$ : 1,3-ddPy$^+$**

(C) A homogeneous mixture of all isomers listed in Table 2

A visual comparison of these mixtures is given in Fig. S7–S9 in the ESI.† In general, all three mixtures show a very comparable and rich spectrum and all of the calculated bands show counterparts in the recorded spectrum except for the strong band in the 721–864 cm$^{-1}$ range. This band, coming exclusively from the **1,3-ddPy$^+$** isomer, is clearly visible in an unweighted mixture of all fourteen isomers as well as the weighted 1 : 1 : 0.5 mixture with the three lowest energy isomers, but is not present in the measured spectrum. Although the 1 : 1 : 0.5 mixture reproduces the experimental spectrum well, suggesting that a 20% contribution of **1,3-ddPy$^+$** is plausible, the 1 : 1 mixture of only **4,5-ddPy$^+$** and **1,2-ddPy$^+$** (Fig. 6) provides a better match to the measured spectrum given the absence of the 721–864 cm$^{-1}$ band.

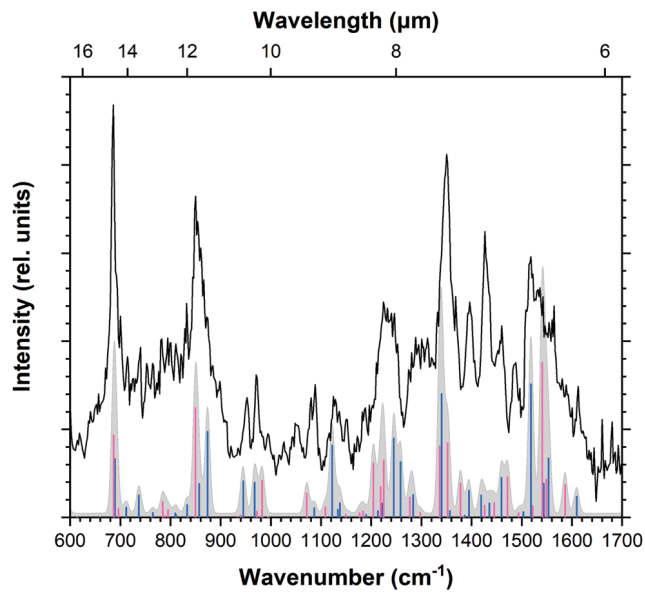

**Fig. 6** IRPD spectrum of ddPy⁺ (black trace) with B3LYP/6-311++G(2d,p) calculated spectra of **4,5-ddPy⁺** (pink sticks) and **1,2-ddPy⁺** (blue sticks) plotted over the Gaussian convolved spectrum (FWHM of 8 cm⁻¹) of the 1 : 1 mixture (A) of both isomers (grey).

Given that all remaining isomers are relatively much higher in energy, it is reasonable to assume that they do not substantially contribute to the overall isomeric mixture.

Close inspection of the spectra reveals that, in mixture A, the relative intensities of the bands between 1200 and 1300 cm⁻¹ corresponding to the C–H in-plane bending modes more closely follow those observed in the recorded spectrum. This mixture has clear counterparts to all the observed bands in the recorded spectrum. Mixtures B and C show a transition present at 1128 cm⁻¹. DFT calculations predict this as a C–H in-plane bending mode for **1,2-ddPy⁺**. This transition is almost entirely imperceptible in mixture C, reflecting the low abundance of this isomer in this mixture. Its detection in the experimental spectrum suggests that, at the very least, **1,2-ddPy⁺** must be a dominant isomer in the ion mixture.

A complementary saturation–depletion measurement was conducted to attempt quantification of branching over the various ddPy⁺ isomers. For this, the relative peak depletion signal is recorded as a function of deposited energy (*i.e.* number of laser pulses applied to the ion cloud) at a wavelength coincident with the centre of an isomer-specific vibrational band. The relative proportion of ion counts at a resonant frequency and an off-resonant frequency reveals the fraction of absorbing ions at this wavelength. The ion counts are determined at the same storing time and laser power to account for the effects of ion losses from the trap by non-radiative processes as well as heating of the ion trap by laser irradiation. The energy can be varied by changing the irradiation time (equivalent to the storage time in the trap, which can be extended to tens of seconds) and/or the light intensity. A similar method has been described elsewhere.[26,27,45]

Guided by scaled harmonic DFT calculated modes, a vibrational band unique to both **4,5-ddPy⁺** and **1,2-ddPy⁺** was chosen as the target for the saturation–depletion study. The resulting scan and its analysis are given in Fig. S10, ESI.† The band at 1245 cm⁻¹ (indicated with a pink arrow in Fig. 3) was predicted to originate from the C–H in-plane modes of both isomers, and showed a maximum saturation of 80%, *i.e.* **4,5-ddPy⁺** and **1,2-ddPy⁺** make up a total fraction of 80% of the ion population (as shown in the ESI†).

An additional scan was then carried out over the C–H out-of-plane mode over 952 cm⁻¹ (marked by a blue arrow in Fig. 3), a band characteristic to the **1,2-ddPy⁺** isomer. This scan was performed using a high laser power (15 mJ per pulse) and trapping the target ions for longer (5.7 s). At this high power, the laser saturates the band completely, allowing the percentage of the isomer that is present in the mixture to be determined. The band was saturated to 40%, revealing the relative proportion of the **1,2-ddPy⁺** isomer. This, in combination with the saturation–depletion study, is consistent with the conclusion that the **4,5-ddPy⁺** and **1,2-ddPy⁺** dominate at equal abundances in the isomeric mixture. This reflects the importance of kinetics in the fragmentation process.

The loss of two hydrogen atoms upon ionisation of pyrene occurs after the migration of one hydrogen atom to its neighbouring carbon, creating an aliphatic CH₂ group at this site.[21,46] Following this, the two hydrogens are lost either sequentially (H + H) or as molecular hydrogen (H₂). The bond dissociation energy for the first hydrogen atom removal was found to be 4.8 eV, regardless of its position.[21] For the second dehydrogenation step, 3.7 eV is required to remove the hydrogen atom from an adjacent position, and 4.8 eV for all other positions. This decrease in energy for the second hydrogen-loss from an adjacent position is due to the spin pairing of the involved carbon atoms, which creates a triple bond to stabilise the lowest energy structure. This complements the results from other studies well.[21,47] More energy is required for the hydrogen to roam to non-adjacent carbon atoms, thus limiting the double dehydrogenation process to form predominantly **4,5-ddPy⁺** and **1,2-ddPy⁺**. This can be qualitatively understood as reflecting the formation of a new bond in these isomers. **1,3-ddPy⁺** can form when the hydrogen at position 1 (Fig. 1) moves to position 2 and then position 3, creating a CH₂ unit which then undergoes double-dehydrogenation. A CH₂ unit is also formed when the hydrogen is at position 2. Dehydrogenation at this stage is energetically more favourable than roaming further to position 3, leading to the formation of **1,2-ddPy⁺** instead. Experiments have also shown that the barrier for a hydrogen atom to roam from one ring to another (*via* a tertiary carbon) is much higher than to roam within a ring.[21] The measured IR spectrum is consistent with these theory-based expectations that the **4,5-ddPy⁺** and **1,2-ddPy⁺** species will dominate the mixture, while the **1,3-ddPy⁺** isomer is at most a minor contributor.

To elucidate the dissociation process and the resulting isomeric distribution, the potential energy surface (PES) connecting the isomers was explored (Fig. 7). The geometries of all transition and intermediate states along with the ωB97XD/6-311++G(2d,p) values of the most stable isomers and the transition states connecting these are available in Fig. S11 and S12, ESI.† It should be noted that following the loss of the second hydrogen atom, ddPy$^+$ may still have enough internal energy to allow for hydrogen-roaming, offering pathways for other isomers to form as well. This scrambling is affected by the barriers involved, with the lowest barrier (between **1,2-ddPy$^+$** and **1,4-ddPy$^+$**) (TS8, Fig. 7) being 2.90 eV at the B3LYP/6-311++G(2d,p) level. This barrier is slightly lower at the wB97XD/6-311++G(2d,p) level (see Table S0, ESI†). All other roaming transitions have much larger barriers as movement of hydrogen from one ring to the next involves bridging a carbon atom: 3.45 eV (TS1, Fig. 7) and 3.86 eV (TS9, Fig. 7). Since the energy of the ionising electrons (20–40 eV) is well above the appearance energy for this system, in principle, all barriers can be easily overcome if all of this energy were transferred.[48] However, the recorded spectra suggest that only 20% of the ion population may exist as isomers other than **4,5-ddPy$^+$** and **1,2-ddPy$^+$**. Therefore we conclude that after the formation of these two isomers, the remaining internal energy in ddPy$^+$ does not allow for efficient hydrogen-roaming.

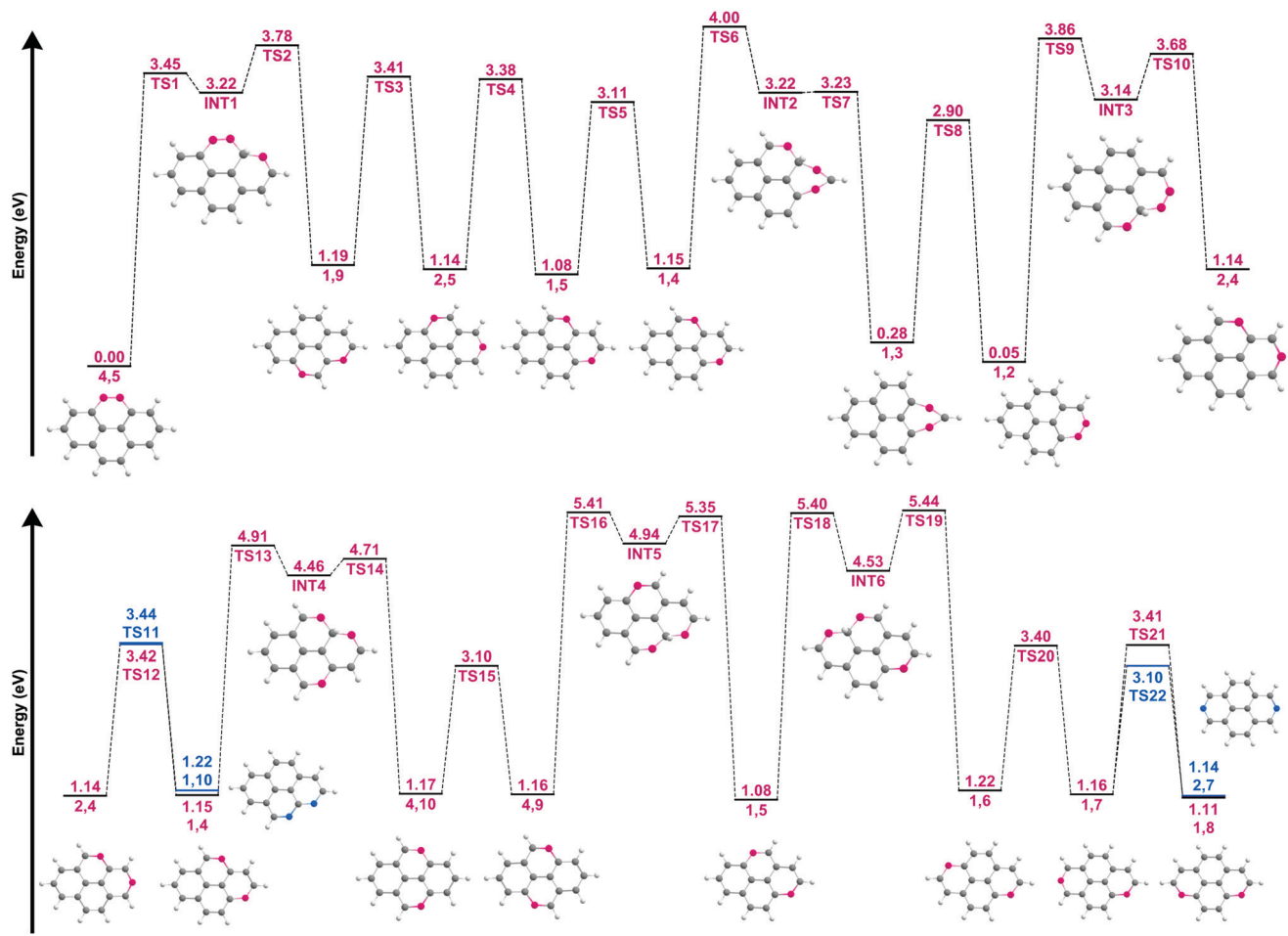

**Fig. 7** Potential energy surface for the hydrogen walk between the fourteen ddPy$^+$ isomers listed in Table 2. The isomers are characterised by their missing hydrogen atoms (shown in pink and blue) using the labelling system shown in Fig. 1. The corresponding structures for the isomers and their intermediate states (INT) are shown. Energies are given with respect to that of the most stable isomer (**4,5-ddPy$^+$**). Transition states are labelled as TS. Blue is used to distinguish pathways to an additional isomer from a single species.

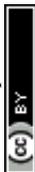

From a spectroscopic perspective, the reduced intensity around the highly characteristic band of **1,3-ddPy$^+$** at 790 cm$^{-1}$ points to a low abundance of this isomer. This, coupled with the energetics of the dissociation process (which favours the loss of directly adjacent hydrogens), is fully consistent with a picture in which the ddPy$^+$ mixture is dominated by **4,5-ddPy$^+$** and **1,2-ddPy$^+$**.

## 4 Astronomical implications

The recorded ddPy$^+$ spectrum extends the existing literature. While there is no strong signature of dehydrogenated PAHs in the studied frequency range of this work, we expect that the characteristic C≡C stretch may allow unambiguous detection of partially dehydrogenated PAHs in space. Unscaled harmonic DFT calculations predict this mode at 2026 cm$^{-1}$ for **4,5-ddPy$^+$** and 1945 cm$^{-1}$ for **1,2-ddPy$^+$**. Future studies would benefit from exploring this mode in further detail.

The present work adds insight on small PAHs involved in the formation of H$_2$ in space. Molecular hydrogen is the most abundant molecule in the interstellar medium. It is generally accepted that, in diffuse clouds, $H_2$ is formed on cold (10–15 K) dust grain surfaces through the recombination of hydrogen atoms in a Langmuir–Hinshelwood mechanism.[49,50] This is supported by laboratory experiments.[51,52] However, $H_2$ is also abundant in regions close to luminous stars known as photo-dissociation regions (PDRs) where the dust temperature is much higher (30–75 K) and the residence time of atomic hydrogen on a grain surface is too short to allow efficient $H_2$ formation.[52] In fact, analysis of the data for PDRs suggests that the $H_2$ formation rate is actually much higher here than in diffuse clouds,[53] suggesting an efficient $H_2$ formation process that is active in regions with abundant UV light. IR observations show that IR emission of UV-pumped PAH species is very bright in these regions and, as reviewed by Wakelam *et al.*,[54] it is therefore likely that PAHs play a role in the $H_2$ formation process. Given the high UV fields in PDRs, $H_2$ formation is thought to be triggered by UV absorption,[7] which can lead to dissociative ionisation with $H_2$ as well as H-loss channels.[21,55,56]

Recent work by Castellanos *et al.* addresses the dehydrogenation processes of a number of small interstellar PAHs in detail.[21,57] They found that PAH size and edge topology influence the preferred hydrogen-loss channel, with smaller PAHs such as pyrene losing hydrogen atoms sequentially (H + H) and larger ones as molecular hydrogen ($H_2$). Their analysis revealed that hydrogen-roaming is an important process in the fragmentation of highly excited PAHs, converting aromatic hydrogen in $sp^2$ sites into $CH_2$ $sp^3$ sites from which either H or $H_2$ can be lost. The formation of aliphatic-like side groups is critical in fragmentation and sets the balance of the competition between H- and $H_2$-loss. Additionally, their analysis revealed that hydrogen-roaming from one ring to an adjacent ring of the PAH is inhibited.[57]

Our study on dehydrogenated pyrene supports this analysis. The IR spectrum of doubly-dehydrogenated pyrene shows that the two hydrogens atoms are lost from the same ring. Hence, roaming from one ring to the next one followed by hydrogen-loss and the creation of two rings each with a radical site is not important. This process of creating an aliphatic $sp^3$ site provides a viable channel for the loss of $H_2$ in an interstellar environment. It is yet to be shown that hydrogen-loss in a concerted fashion from PAHs can be important in PDRs. An interesting follow-up study would be to explore the doubly-dehydrogenated states of astronomically-sized PAHs ($\simeq 50$ carbon atoms), and to further investigate the effects of edge topology on hydrogen loss to gain an insight into the photoproducts of PAHs in regions with high UV radiation. Of course, the study of such larger PAHs would come with additional challenges, the foremost being an even greater number of possible isomers with energetically low-lying geometries.

# 5 Conclusions

The IR spectrum of the pyrene radical cation has been recorded using the neon-tagging pre-dissociation technique. The measured band positions are shown to compare well with the anharmonic calculations. The spectrum of $Py^+$ presented here adds spectral information and accuracy to existing data. All bands having DFT-predicted intensities larger than 10 km mol$^{-1}$ have now been detected in the gas phase. With the application of a suitable scaling factor to account for anharmonicity, the overall profile of the convolved harmonic spectrum is similar to that of the anharmonic one, and the scaled harmonic calculations are thus shown to provide a good benchmark for the isomeric assignments in the ddPy$^+$ spectrum. The good fit between the experimental and calculated spectra indicates that the influence of the neon-tag need not be calculated as it is negligible for these larger systems.

The first experimental IR spectrum of ddPy$^+$ is presented. Based on a comparison with the scaled harmonic DFT calculations shown here, the recorded spectrum cannot be reproduced with only one isomer but a fair reproduction is realised by combining the spectral features of two isomers: **4,5-ddPy$^+$** and **1,2-ddPy$^+$**. This is supported by depletion studies that reveal that the spectra are dominated at the 80% level by these two isomers. In addition, there is no evidence for the presence of the very strong mode of **1,3-ddPy$^+$**, the next most stable isomer, in the 721–864 cm$^{-1}$ range. Theoretical calculations support that hydrogen-loss from the same aromatic ring is favoured over hydrogen-loss from separate rings. The dehydrogenation process at a trio-site is initiated by roaming of a hydrogen to form a $CH_2$ group at either position 2 or 3 (Fig. 1), resulting in **1,2-ddPy$^+$** rather than the energetically equally favoured **1,3-ddPy$^+$**. Formation of the latter cannot be fully excluded but likely will not contribute to the overall signal by more than 20%. The data show no unique signatures that allow clear comparison with astronomical IR data on PAH sources, but allow to conclude, in agreement with earlier work, that the dissociative ionisation of a small PAH like pyrene may contribute to interstellar $H_2$ formation.

The work presented here illustrates a number of advantages of neon-tagging IRPD spectroscopy. It is possible to record high resolution gas-phase spectra of ions that are difficult to study using alternate methods, particularly at low molecular abundances. However, this experimental method is mass-selective and not isomer-selective. As a consequence, the spectroscopic investigation of fragment species is only possible with complementary theoretical data, as shown here. Similar studies on larger PAHs will be very useful, linking the structural findings derived for pyrene to astronomically-relevant PAHs. Future progress in such fragmentation studies is reliant on the combined advancement of highly sensitive instrumentation and theoretical quantum chemistry in computing anharmonic spectra of large aromatic molecules.

# Conflicts of interest

There are no conflicts to declare.

# Acknowledgements

SP and SBa acknowledge the European Union and Horizon 2020 funding awarded under the Marie Skłodowska-Curie

action to the EUROPAH consortium (grant number 722346). JB acknowledges the Netherlands Organisation for Scientific Research (Nederlandse Organisatie voor Wetenschappelijk Onderzoek, NWO) for a VIDI grant (grant number 723.016.006). AC acknowledges NWO for a VENI grant (grant number 639.041.543). This work was supported by NWO Exact and Natural Sciences for the use of supercomputer facilities (grant numbers 16638, 17676, and SH-362-15). We thank Aravindh Nivas Marimuthu for support with the data analysis, and Olivier Burggraaff for helping create Fig. 7. The authors greatly appreciate the experimental support provided by the FELIX team, and acknowledge the NWO for the support of the FELIX Laboratory. We thank the Cologne Laboratory Astrophysics group for providing the FELion ion trap instrument for the current experiments and the Cologne Center for Terahertz Spectroscopy (core facility, Deutsche Forschungsgemeinschaft grant SCHL 341/15-1) for supporting its operation. We thank the two anonymous reviewers for their thorough and constructive reviews.

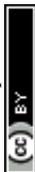

**Electronic Supporting Information (ESI)**

**Structural investigation of doubly-dehydrogenated pyrene cations**

Sanjana Panchagnula,*[a,b] Jordy Bouwman,[a] Daniël B. Rap,[c] Pablo Castellanos,[a] Alessandra Candian,[a,d] Cameron Mackie,[a] Shreyak Banhatti,[e] Sandra Brünken,[c] Harold Linnartz,[a] and Alexander G.G.M. Tielens[b]

[a] *Laboratory for Astrophysics, Leiden Observatory, Leiden University, PO Box 9513, 2300 RA Leiden, The Netherlands.*

[b] *Leiden Observatory, Leiden University, PO Box 9513, 2300 RA Leiden, The Netherlands.*

[c] *Radboud University, Institute for Molecules and Materials, FELIX Laboratory, Toernooiveld 7, NL-6525ED Nijmegen, The Netherlands.*

[d] *Van 't Hoff Institute for Molecular Sciences, University of Amsterdam, PO Box 94157, 1090 GD Amsterdam, The Netherlands.*

[e] *I. Physikalisches Institut, Universität zu Köln, Zülpicher Str. 77, 50937 Köln, Germany.*

*To whom correspondence should be addressed.

**E-mail:** panchagnula@strw.leidenuniv.nl

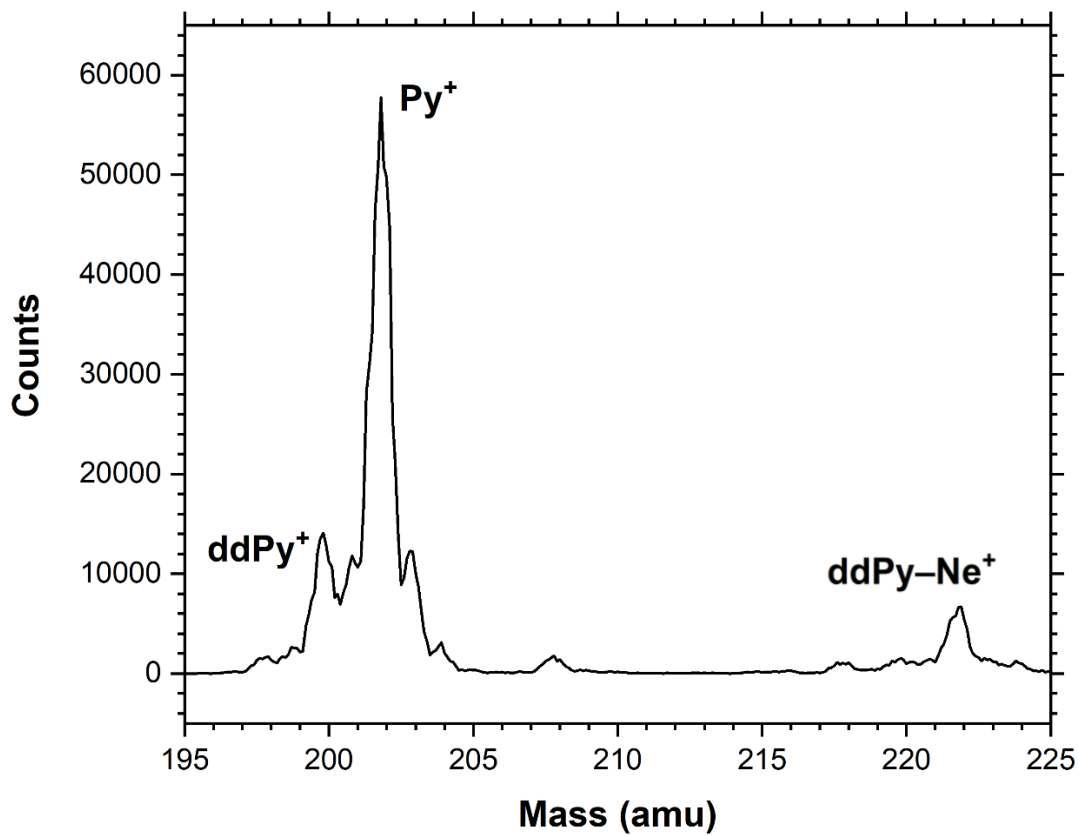

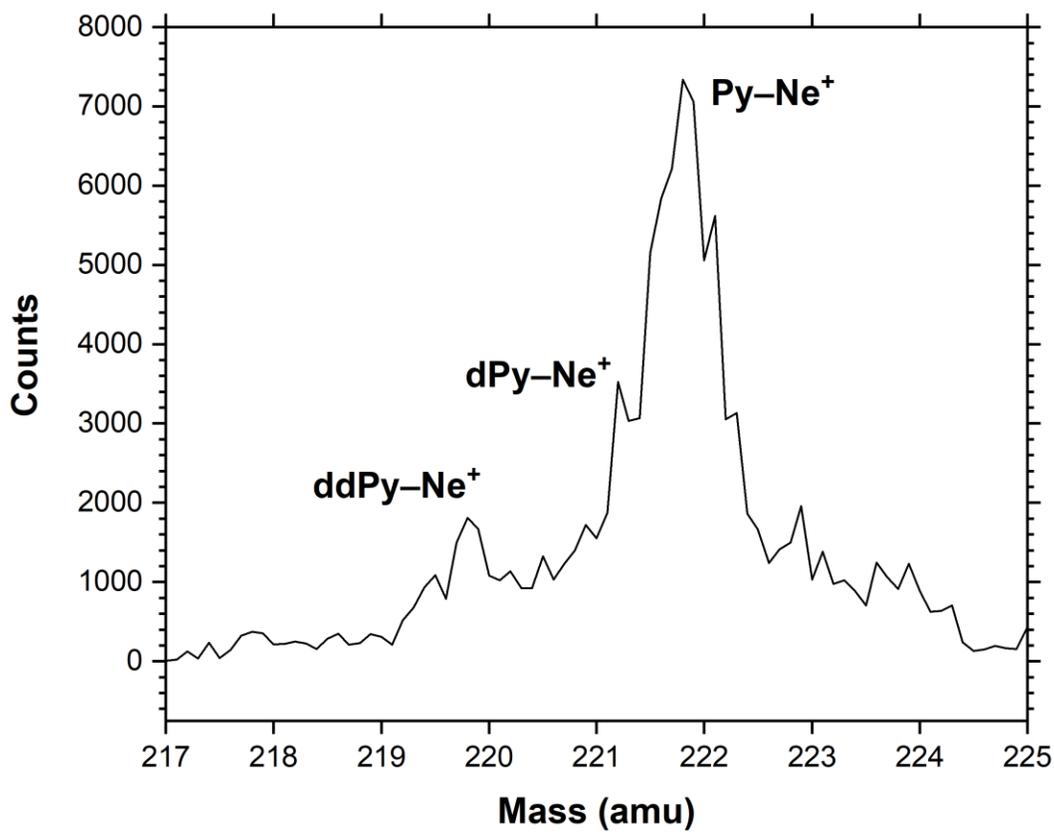

**Fig. S1** Mass spectrum of pyrene showing Py⁺ at 202 amu, ddPy⁺ at 200 amu, ddPy–Ne⁺ at 220 amu, dPy–Ne⁺ at 221 amu, and Py–Ne⁺ at 222 amu.

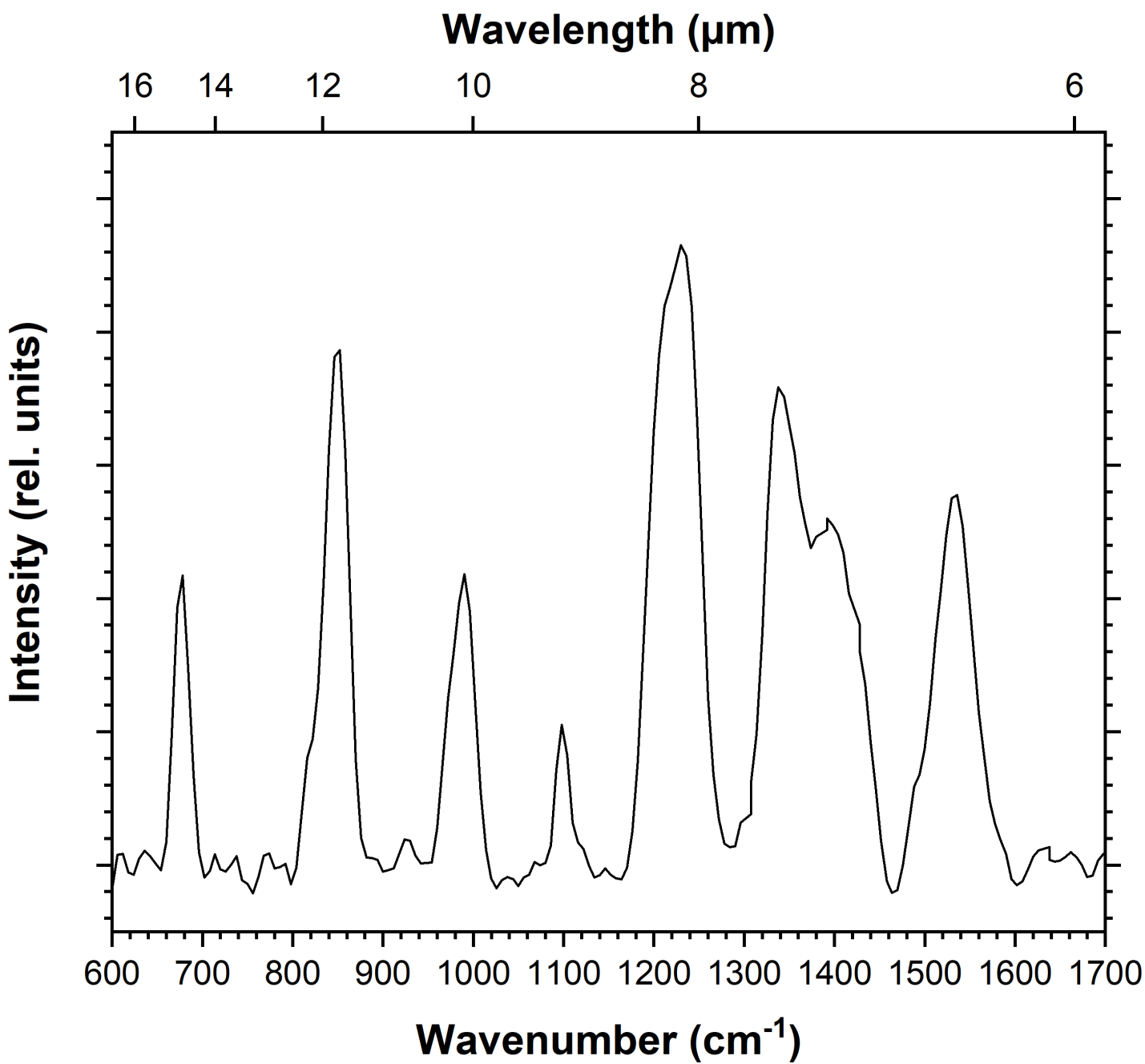

**Fig. S2** IRMPD spectrum of Py⁺ using FELIX and a Paul-type quadrupole ion trap (unpublished data). See Bouwman *et al.* for experimental details.[58]

The recorded IRMPD spectrum shows improvement in band resolution and strength in the 600–1300 cm⁻¹ range compared to earlier IRMPD data presented in the literature,[19] but clearly lacks resolution compared to the Ne–tag experiments presented in this work.

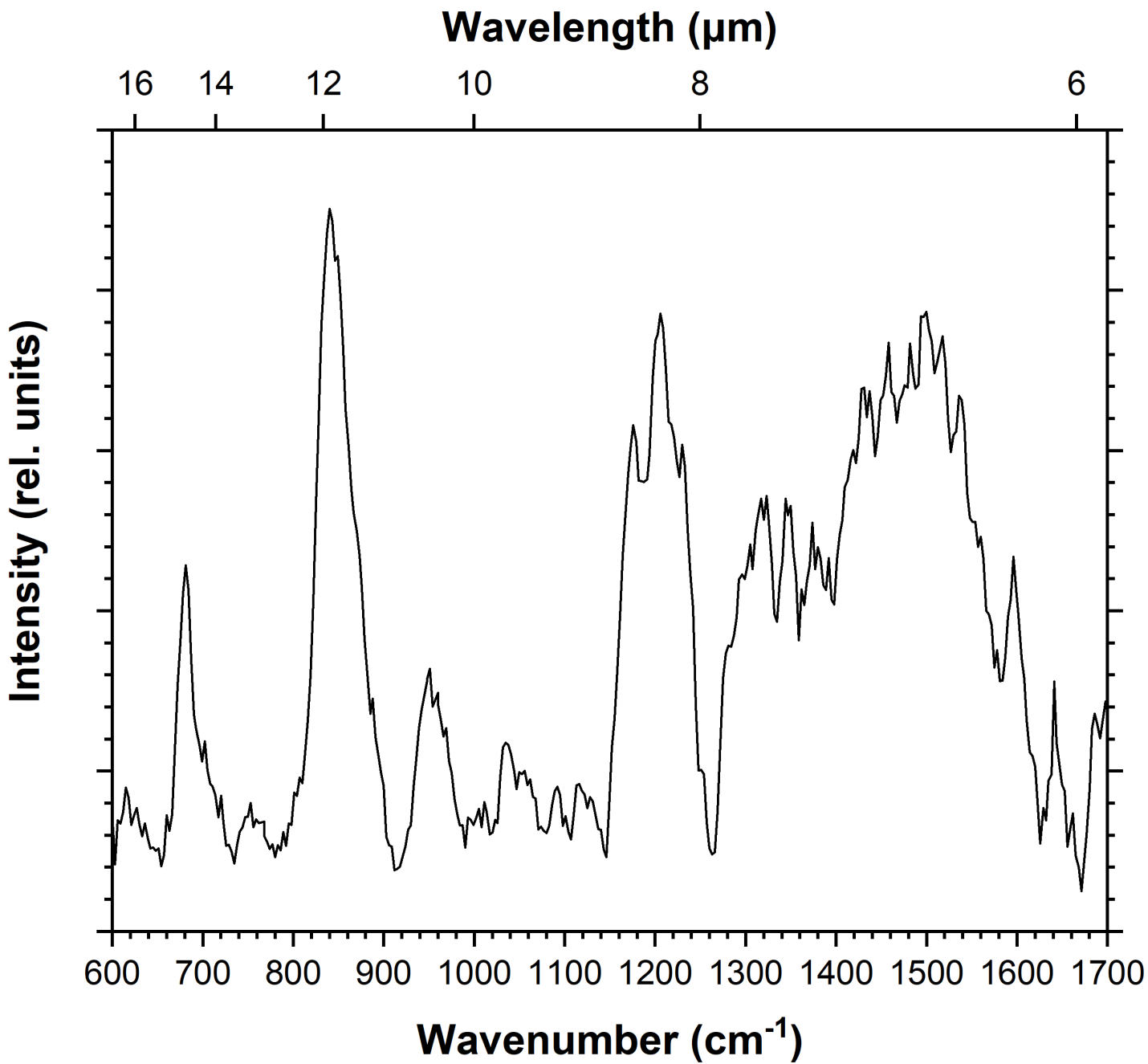

**Fig. S3** IRMPD spectrum of ddPy⁺ using FELIX and a Paul-type quadrupole ion trap (unpublished data).

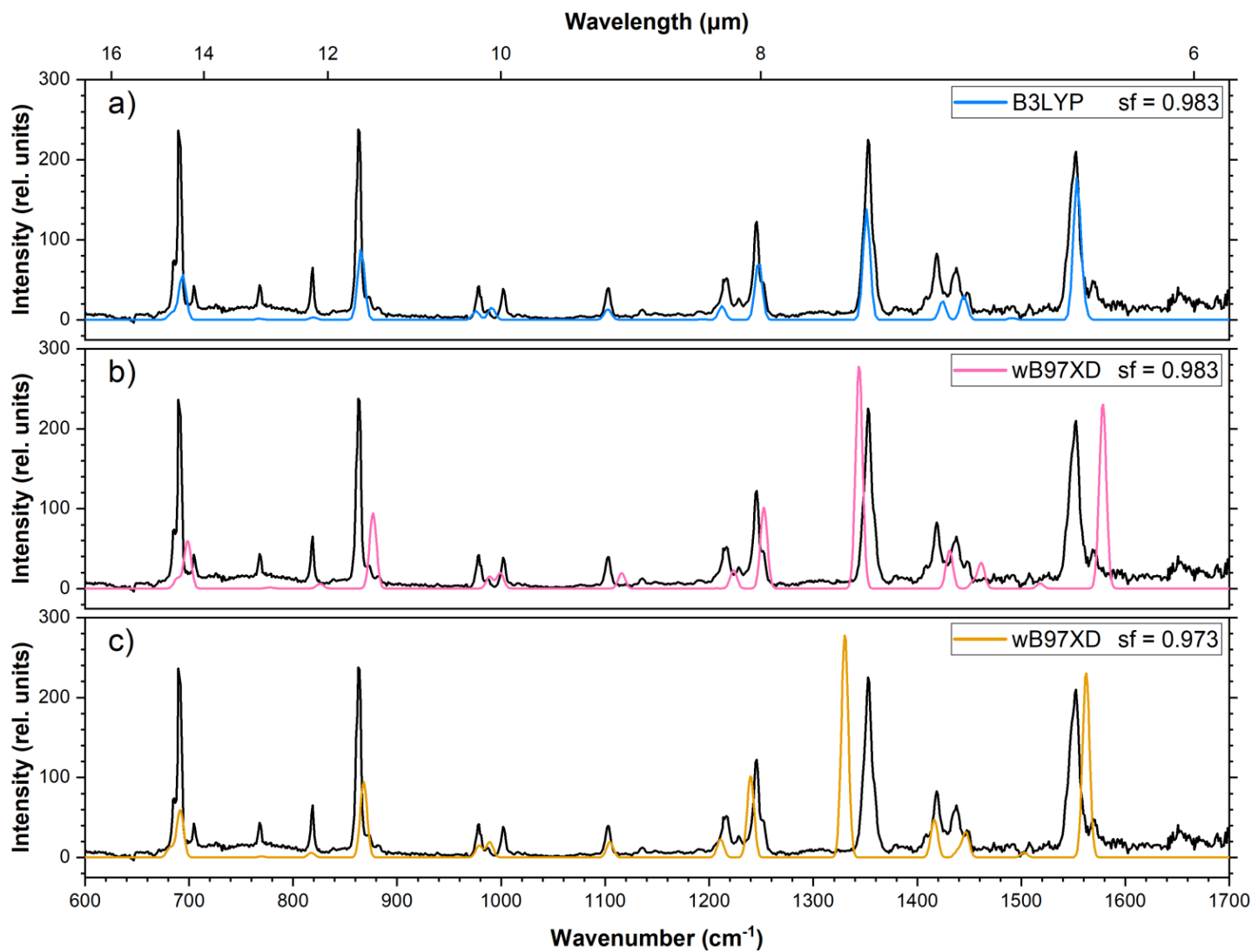

**Fig. S4** IRPD spectrum of Py⁺ (black) compared to computationally generated spectra using (a) B3LYP (blue, scaled by 0.983), (b) wB97XD (pink, scaled by 0.983), and (c) wB97XD (orange, scaled by 0.973). All functionals used the 6-311++G(2d,p) basis set.

For wB97XD, a scaling factor of 0.973 serves as a compromise between the C=C stretch region (~1500 cm⁻¹) and the bending modes (~700-1000 cm⁻¹) to achieve the best possible fit. In agreement with the literature, B3LYP/6-311+G(2d,p) is much preferred in vibrational analysis over wB97XD/6-311++G(2d,p) when scaled by a factor of 0.983 (blue trace, Fig. R1). Nevertheless, we take the empirically-determined scaling factor of 0.973 and apply it to our ddPy⁺ calculations with wB97XD.

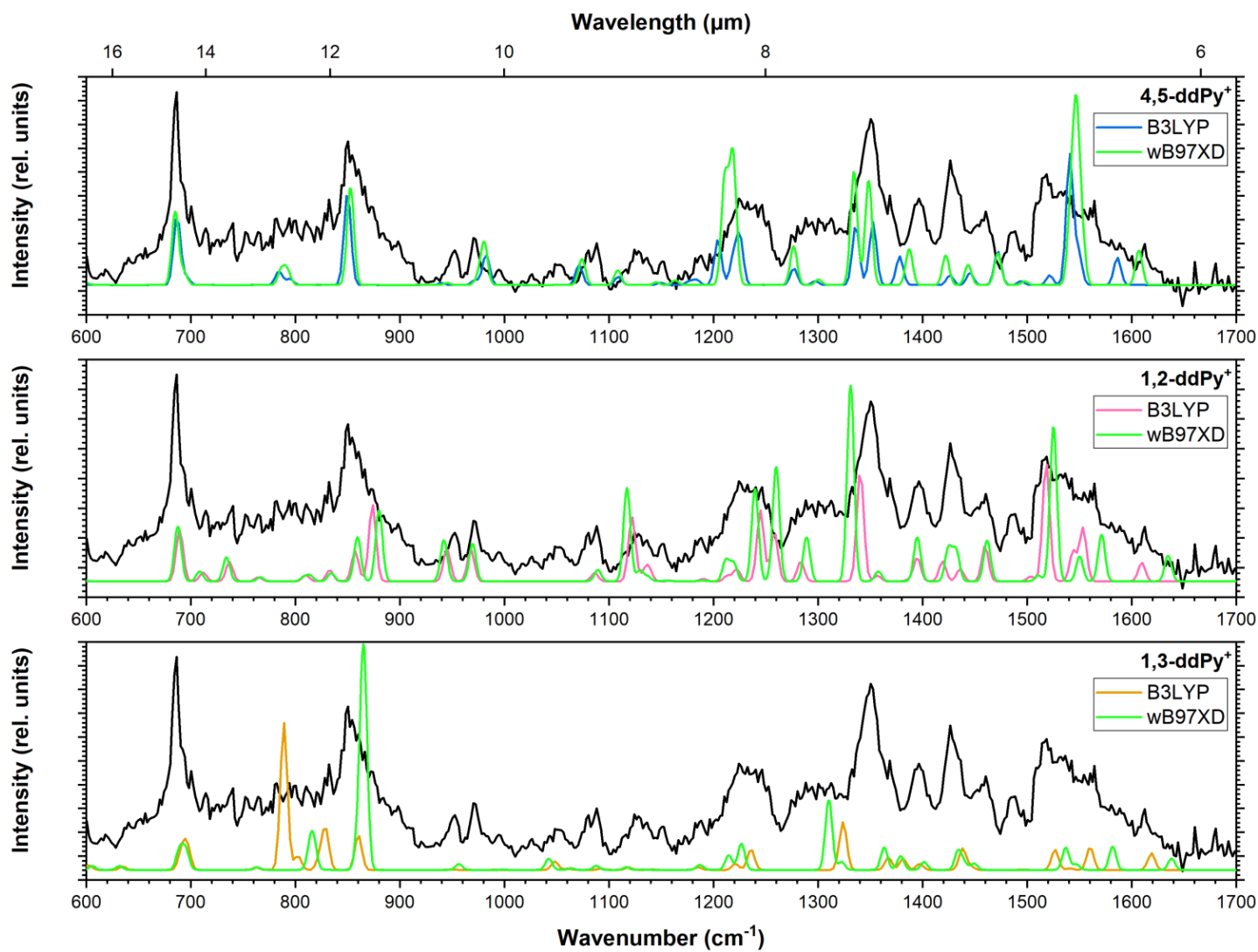

**Fig. S5** IRPD spectrum of ddPy$^+$ (black) compared to computationally generated spectra of (a) 1,2–ddPy$^+$, (b) 4,5–ddPy$^+$, and (c) 1,3–ddPy$^+$. Calculations were performed using B3LYP (scaled by 0.983) and wB97XD (scaled by 0.973) with the 6-311++G(2d,p) basis set.

While there are some minor differences in band positions and intensities for 4,5–ddPy$^+$ and 1,2–ddPy$^+$, it is the cyclopropenyl transition in 1,3–ddPy$^+$ that is the most strikingly dissimilar between the two computational methods.

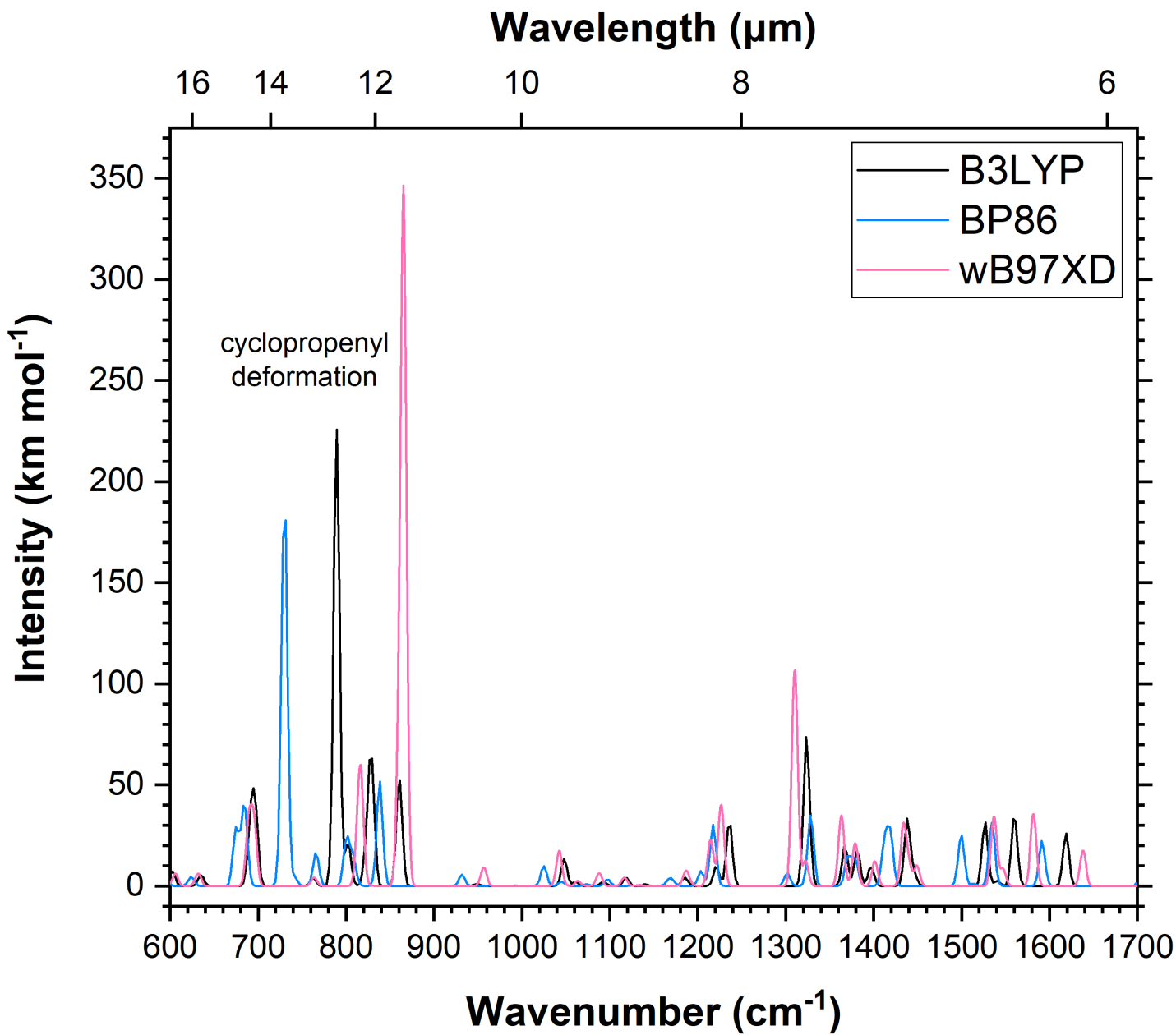

**Fig. S6** Theoretical IR spectra for 1,3–ddPy⁺ using B3LYP (black), BP86 (blue), and wB97XD (pink). All calculations were performed with the 6-311++G(2d,p) basis set.

With wB97XD, the cyclopropenyl transition appears at 864 cm⁻¹ even more intensely than for B3LYP. BP86 predicts this band at 721 cm⁻¹. This shows that the modelling of the cyclopropenyl vibrational mode is rather sensitive to the choice of computational method (with a shift of 69 cm⁻¹ in B3LYP vs. BP86, and 74 cm⁻¹ in B3LYP vs. wB97XD).

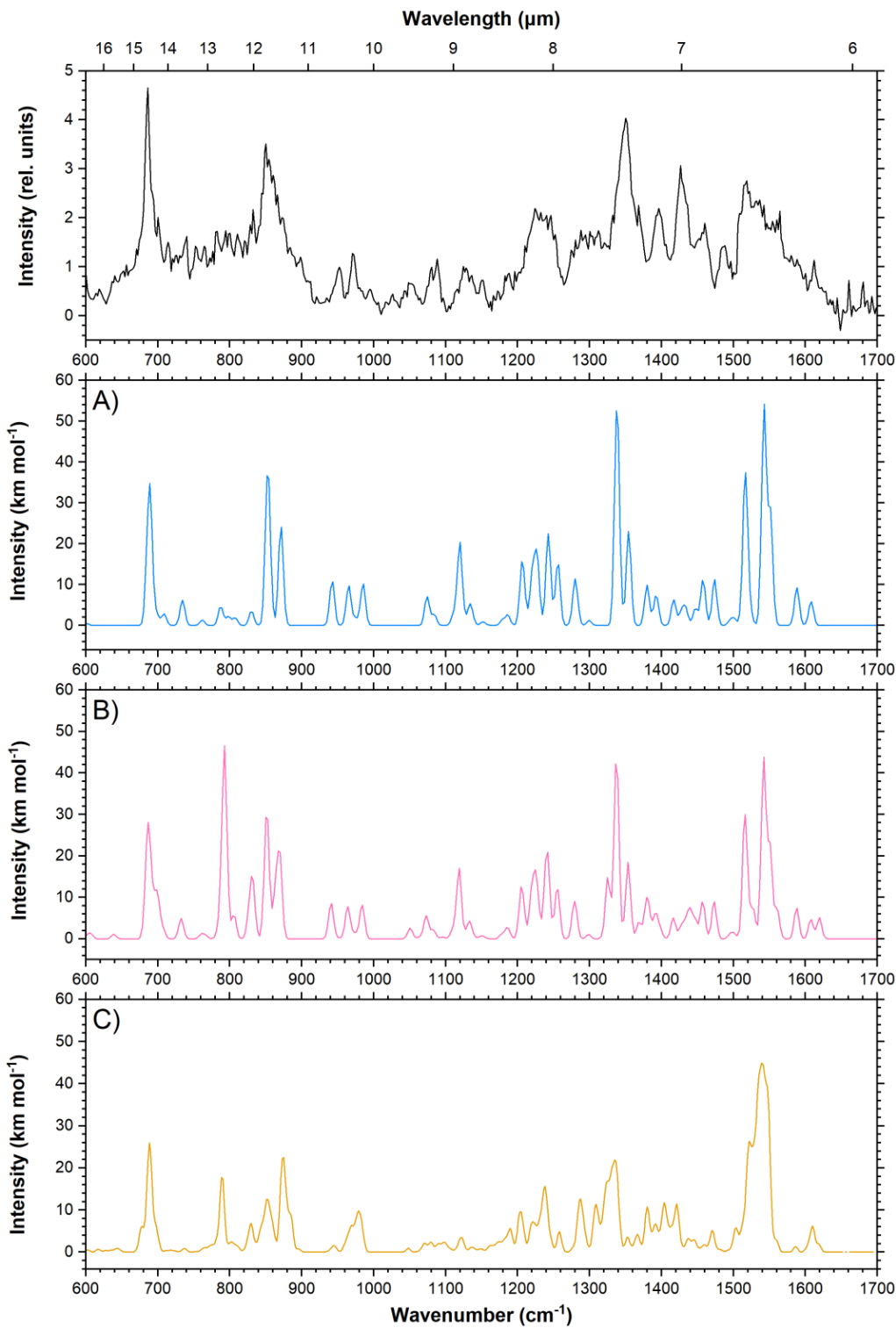

**Fig. S7** IRPD spectrum of ddPy⁺ (black) compared to computationally generated spectra using the B3LYP/6-311++G(2d,p) calculated spectra of the possible isomers (scaled by 0.983, convolved using a Gaussian profile with a FWHM of 8 cm⁻¹). Pink shows a 1:1 mixture (A) of 4,5–ddPy⁺ and 1,2–ddPy⁺. Blue shows a 1:1:0.5 mixture (B) of 4,5–ddPy⁺ : 1,2–ddPy⁺ : 1,3–ddPy⁺. Green shows a homogeneous mixture (C) of all fourteen isomers.

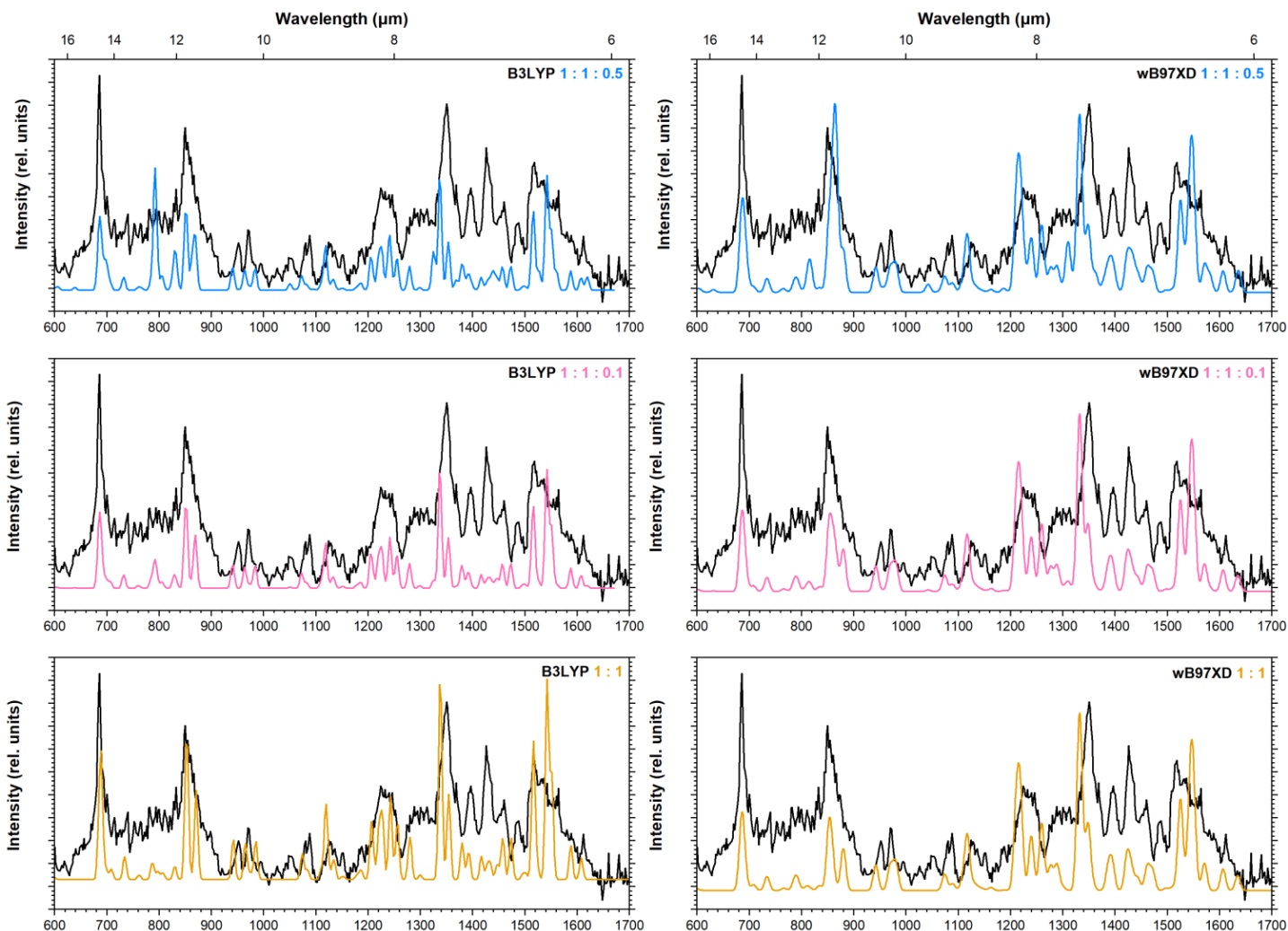

**Fig. S8** IRPD spectrum of ddPy⁺ (black) compared to computationally generated spectra with mixtures of 4,5–ddPy⁺, 1,2–ddPy⁺, and 1,3–ddPy⁺. Left panels show B3LYP/6-311++G(2d,p) spectra scaled by 0.983, while right panels show wB97XD/6-311++G(2d,p) spectra scaled by 0.973. Blue spectra show a 1:1:0.5 mixture of all three isomers. Pink spectra show a 1:1:0.1 mixture of all three isomers. Orange spectra show a 1:1 mixture of only 4,5–ddPy⁺ and 1,2–ddPy⁺.

We see that the calculated spectra match better with the experimentally recorded ddPy⁺ spectrum with diminishing quantities of 1,3–ddPy⁺.

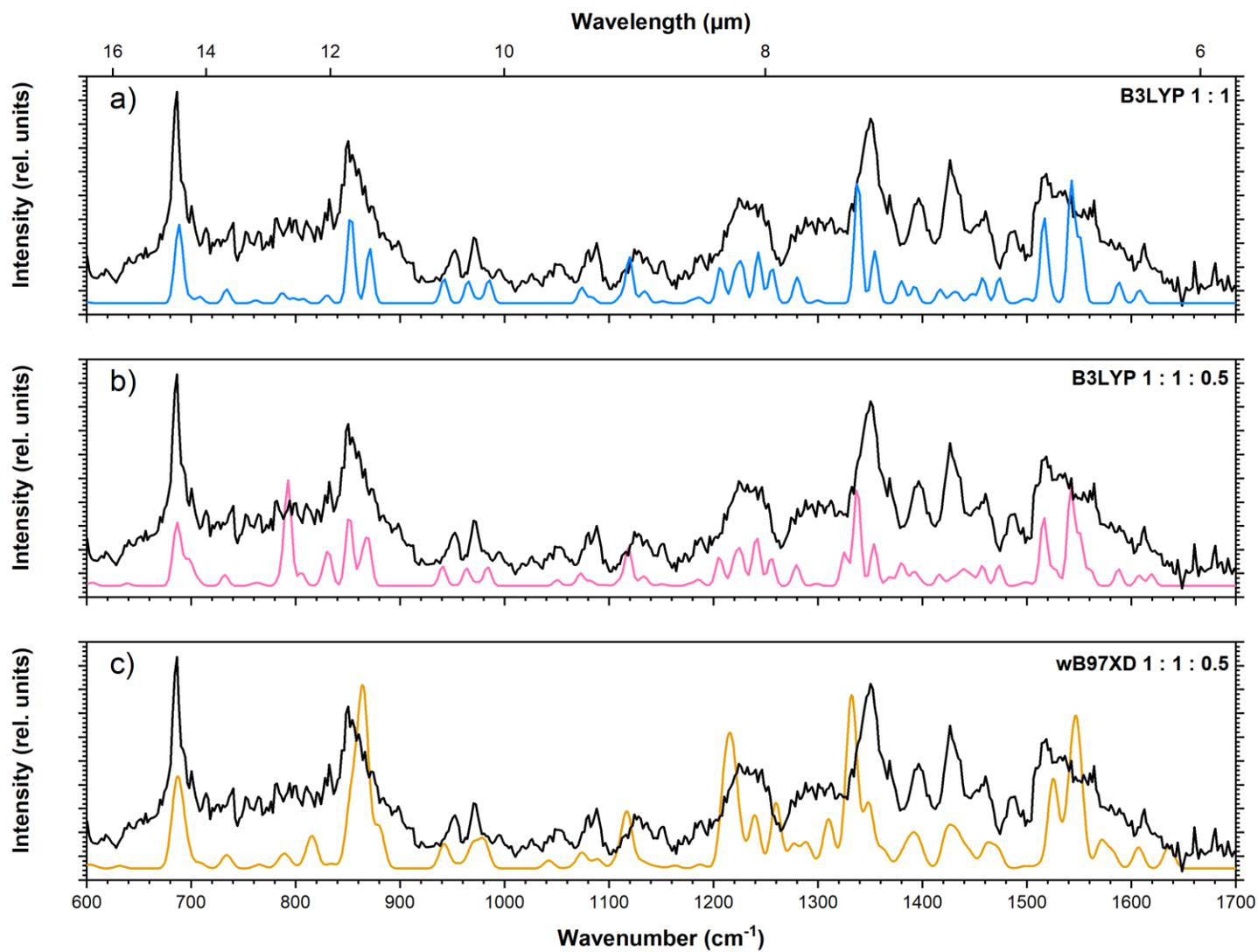

**Fig. S9** IRPD spectrum of ddPy⁺ (black) compared to computationally generated spectra of (a, b) B3LYP (scaled by 0.983), and (c) wB97XD (scaled by 0.973). Blue shows a 1:1 mixture of 4,5–ddPy⁺ and 1,2–ddPy⁺. Pink and orange show a 1:1:0.5 mixture of 4,5–ddPy⁺ : 1,2–ddPy⁺ : 1,3–ddPy⁺. All calculations were performed with the 6-311++G(2d,p) basis set.

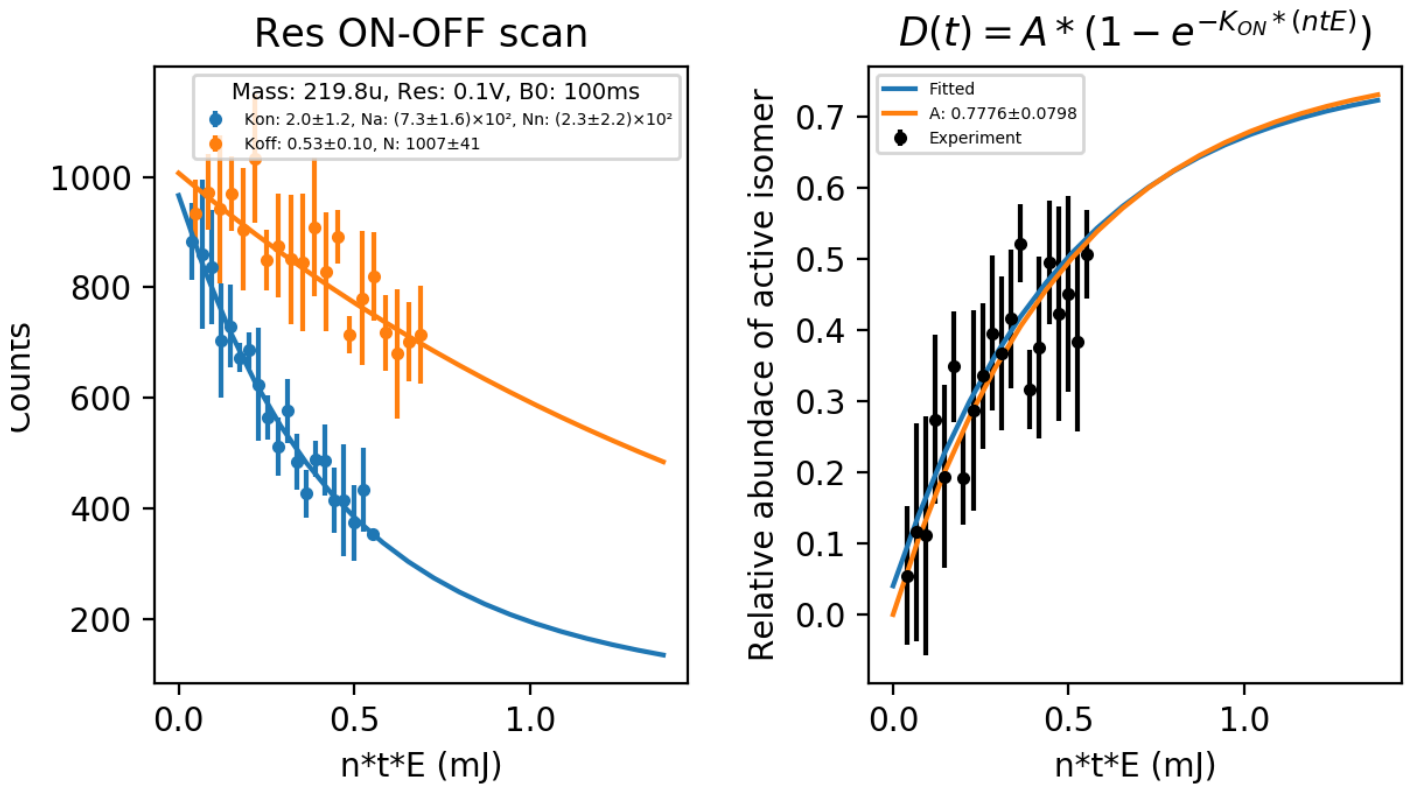

**Fig. S10** Depletion scan of the vibrational band at 1245 cm⁻¹ in the ddPy⁺ spectrum. **Left:** Blue shows the exponential decay of the complex counts ($N_{on}(ntE) = N_a\,e^{-(K_{off}+K_{on})\,ntE} + N_n\,e^{-(K_{off})\,ntE}$) on resonance as a function of total deposited energy, with $t$ trapping time (in s), $E$ energy (in mJ) per macropulse at $n = 10$ Hz rate, $N_a$ total number of active isomers, $N_n$ total number of inactive isomers at $t = 0$ s, and $K_{on}$ and $K_{off}$, the on- and off-resonance decay rate constants. Orange shows the exponential decay of the complex counts off-resonance ($N_{off} = (N_a + N_n)\,e^{-(K_{off})\,ntE}$) due to other loss mechanisms with the rate constant $K_{off}$. **Right:** The relative depletion $D = 1 - N_{on}/N_{off}$, which has been fitted with an exponential function $D(ntE) = A(1 - e^{-(K_{on})\,ntE})$, where $A$ is the constant for relative abundance of the active isomer $A = \frac{N_a}{N_a + N_n}$.

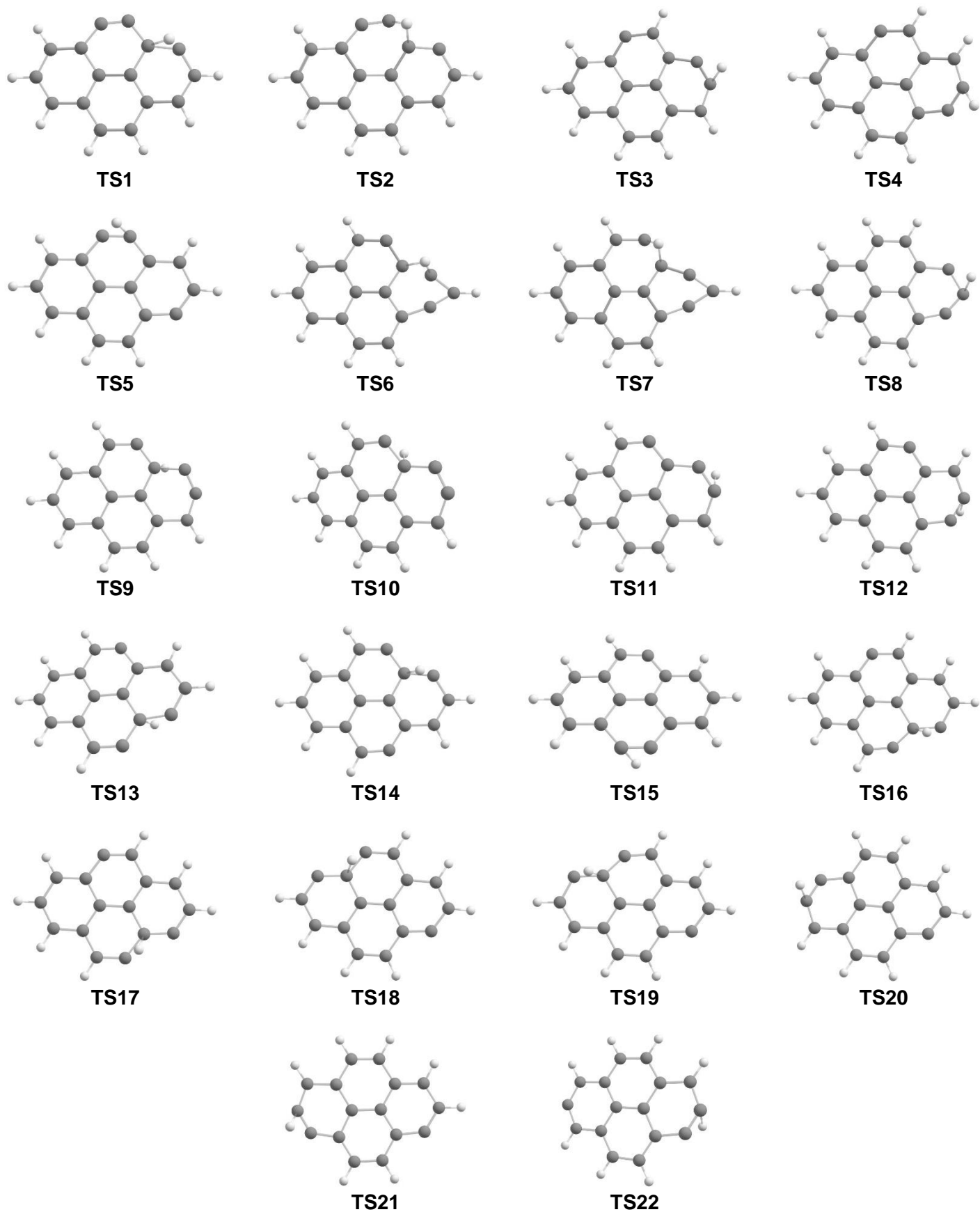

**Fig. S11** Chemical structures of transition states in hydrogen migration pathways.

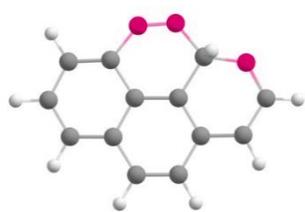
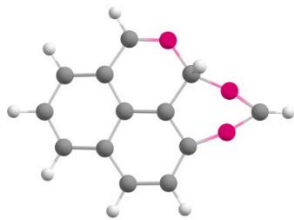
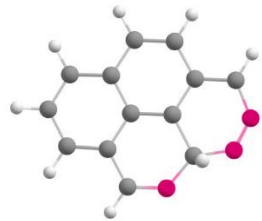

**INT1**   **INT2**   **INT3**

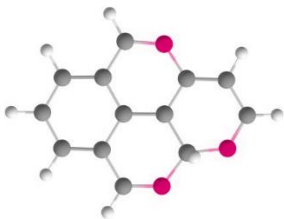
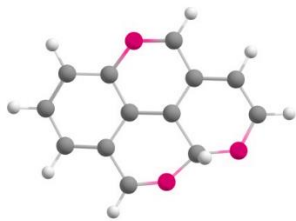
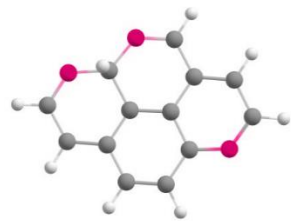

**INT4**   **INT5**   **INT6**

**Fig. S12** Chemical structures of intermediates in hydrogen migration pathways.

**Table S0** DFT calculated energies ($\Delta E$, given in eV) of selected isomers and transition states (shown in Fig. 7 in the manuscript). Energies presented are relative to 4,5–ddPy$^+$.

| | $\Delta E$ (eV) | |
| --- | --- | --- |
| **Isomer** | **B3LYP** | **wB97XD** |
| **4,5–ddPy$^+$** | 0.00 | 0.00 |
| **1,2–ddPy$^+$** | 0.05 | 0.04 |
| **1,3–ddPy$^+$** | 0.28 | 0.04 |
| **TS1** | 3.45 | 3.36 |
| **TS8** | 2.90 | 2.84 |
| **TS9** | 3.84 | 3.86 |

| Atomic number | Coordinates (Å) | | |
|---|---|---|---|
| | X | Y | Z |
| 6 | −0.005 003 | 0.621 296 | −2.509 953 |
| 6 | 0.010 855 | −0.595 086 | −2.404 226 |
| 6 | 0.004 561 | −1.500 971 | −1.323 322 |
| 6 | −0.031 210 | −0.721 300 | −0.095 388 |
| 6 | −0.050 843 | 0.696 648 | −0.087 362 |
| 6 | −0.086 898 | 1.474 240 | −1.368 204 |
| 6 | −0.017 168 | 1.413 845 | 1.134 165 |
| 6 | −0.012 805 | 0.685 802 | 2.361 645 |
| 6 | −0.024 520 | −0.674 659 | 2.362 040 |
| 6 | −0.026 108 | −1.421 285 | 1.146 156 |
| 6 | 0.026 301 | −2.880 991 | −1.284 487 |
| 6 | 0.016 395 | −3.538 027 | −0.048 637 |
| 6 | −0.006 424 | −2.828 087 | 1.137 174 |
| 6 | 0.020 342 | 2.887 443 | −1.227 645 |
| 6 | 0.039 143 | 3.583 346 | −0.032 919 |
| 6 | 0.031 189 | 2.832 248 | 1.130 181 |
| 1 | 0.052 054 | −3.445 130 | −2.207 400 |
| 1 | 0.032 439 | −4.620 143 | −0.024 854 |
| 1 | −0.003 049 | −3.357 019 | 2.082 473 |
| 1 | −0.021 234 | −1.216 980 | 3.299 962 |
| 1 | 0.002 460 | 1.235 779 | 3.294 191 |
| 1 | 0.070 938 | 4.665 469 | −0.012 971 |
| 1 | 0.058 109 | 3.350 088 | 2.082 818 |
| 1 | −1.161 362 | 2.104 244 | −1.398 549 |

Table 1: B3LYP/6-311++G(2d,p) optimised structure of transition state **TS1**.

| Atomic number | Coordinates (Å) | | |
|---|---|---|---|
| | X | Y | Z |
| 6 | −0.014 620 | 0.614 626 | −2.523 821 |
| 6 | −0.018 154 | −0.614 474 | −2.361 863 |
| 6 | −0.011 796 | −1.521 948 | −1.315 244 |
| 6 | −0.026 453 | −0.724 566 | −0.080 481 |
| 6 | −0.044 603 | 0.689 688 | −0.075 947 |
| 6 | −0.074 033 | 1.478 901 | −1.330 771 |
| 6 | −0.022 716 | 1.410 272 | 1.150 752 |
| 6 | −0.012 131 | 0.672 055 | 2.379 727 |
| 6 | −0.007 288 | −0.683 096 | 2.381 409 |
| 6 | −0.008 962 | −1.432 033 | 1.160 399 |
| 6 | 0.014 129 | −2.910 453 | −1.277 690 |
| 6 | 0.023 951 | −3.552 493 | −0.045 844 |
| 6 | 0.013 879 | −2.831 662 | 1.143 686 |
| 6 | −0.018 133 | 2.869 658 | −1.222 087 |
| 6 | 0.019 723 | 3.560 400 | −0.053 752 |
| 6 | 0.008 137 | 2.813 667 | 1.136 120 |
| 1 | 0.021 666 | −3.473 140 | −2.201 298 |
| 1 | 0.041 165 | −4.634 318 | −0.011 585 |
| 1 | 0.026 529 | −3.363 135 | 2.087 626 |
| 1 | 0.004 812 | −1.228 797 | 3.316 997 |
| 1 | −0.003 194 | 1.222 210 | 3.312 459 |
| 1 | 0.061 184 | 4.642 812 | −0.026 754 |
| 1 | 0.029 589 | 3.343 496 | 2.081 914 |
| 1 | −1.117 387 | 1.224 360 | −2.016 737 |

Table 2: B3LYP/6-311++G(2d,p) optimised structure of transition state **TS2**.

| Atomic number | Coordinates (Å) | | |
|---|---|---|---|
| | X | Y | Z |
| 6 | 2.643 269 | −0.481 661 | 0.020 455 |
| 6 | 0.018 973 | 0.673 282 | −0.000 752 |
| 6 | 2.418 952 | 0.847 435 | 0.023 939 |
| 6 | 0.147 736 | −0.740 749 | 0.013 163 |
| 6 | 1.185 376 | 1.508 980 | 0.010 289 |
| 6 | 1.486 061 | −1.294 332 | 0.009 771 |
| 6 | −1.056 698 | −1.565 916 | 0.016 351 |
| 6 | 1.041 040 | 2.908 188 | 0.006 881 |
| 6 | 1.376 576 | −2.700 224 | −0.077 338 |
| 6 | −2.333 541 | −0.908 862 | 0.015 900 |
| 6 | −0.219 127 | 3.475 302 | −0.004 811 |
| 6 | −2.429 525 | 0.441 919 | −0.001 918 |
| 6 | −1.361 883 | 2.673 062 | −0.010 439 |
| 6 | −1.268 937 | 1.281 960 | −0.008 380 |
| 1 | −3.406 222 | 0.911 600 | −0.005 868 |
| 1 | −2.341 497 | 3.135 969 | −0.015 950 |
| 1 | −3.224 959 | −1.522 689 | 0.023 199 |
| 1 | −0.323 317 | 4.552 773 | −0.008 195 |
| 1 | 1.927 914 | 3.528 234 | 0.013 719 |
| 6 | −1.014 332 | −2.976 088 | 0.008 091 |
| 6 | 0.320 896 | −3.386 416 | −0.027 394 |
| 1 | 3.631 863 | −0.923 873 | 0.011 429 |
| 1 | 1.039 848 | −3.430 105 | 1.016 141 |
| 1 | −1.885 524 | −3.609 894 | 0.051 291 |

Table 3: B3LYP/6-311++G(2d,p) optimised structure of transition state **TS3**.

| Atomic number | Coordinates (Å) | | |
|---|---|---|---|
| | X | Y | Z |
| 6 | 2.543 953 | −0.622 300 | 0.015 973 |
| 6 | −0.020 163 | 0.656 475 | −0.003 458 |
| 6 | 2.381 124 | 0.700 995 | −0.004 182 |
| 6 | 0.050 995 | −0.761 127 | 0.009 696 |
| 6 | 1.180 221 | 1.440 356 | −0.011 042 |
| 6 | 1.349 590 | −1.438 785 | 0.016 150 |
| 6 | −1.206 835 | −1.470 179 | 0.003 971 |
| 6 | 1.100 803 | 2.830 129 | −0.012 388 |
| 6 | 1.478 720 | −2.840 788 | 0.011 770 |
| 6 | −2.449 716 | −0.811 170 | 0.015 474 |
| 6 | −0.142 318 | 3.466 314 | −0.007 346 |
| 6 | −2.479 222 | 0.555 231 | 0.020 259 |
| 6 | −1.311 284 | 2.729 845 | 0.003 218 |
| 6 | −1.281 957 | 1.321 530 | 0.006 958 |
| 1 | −3.429 907 | 1.074 625 | 0.024 222 |
| 1 | −2.270 052 | 3.234 008 | 0.009 193 |
| 6 | −3.361 193 | −1.393 407 | 0.010 128 |
| 6 | −0.187 343 | 4.547 847 | −0.010 508 |
| 1 | 2.013 558 | 3.411 893 | −0.017 084 |
| 1 | −0.927 573 | −2.853 137 | −0.082 131 |
| 1 | 0.201 487 | −3.410 469 | −0.030 828 |
| 1 | 3.516 523 | −1.100 024 | 0.024 142 |
| 1 | 2.419 809 | −3.365 059 | 0.058 356 |
| 1 | −0.515 796 | −3.546 232 | 1.008 106 |

Table 4: B3LYP/6-311++G(2d,p) optimised structure of transition state **TS4**.

| Atomic number | Coordinates (Å) | | |
| --- | --- | --- | --- |
| | X | Y | Z |
| 6 | −2.436 737 | 0.631 261 | −0.010 540 |
| 6 | 0.043 762 | −0.735 985 | 0.015 107 |
| 6 | −2.430 403 | −0.723 115 | −0.010 819 |
| 6 | 0.025 806 | 0.699 271 | 0.015 875 |
| 6 | −1.194 783 | −1.447 883 | −0.000 466 |
| 6 | −1.218 622 | 1.387 509 | 0.000 163 |
| 6 | 1.225 282 | 1.520 453 | 0.014 401 |
| 6 | −1.095 171 | −2.830 306 | −0.001 259 |
| 6 | −1.223 424 | 2.794 620 | −0.001 125 |
| 6 | 2.343 831 | 0.629 022 | −0.013 843 |
| 6 | 0.031 812 | −3.598 165 | 0.002 058 |
| 6 | 2.359 237 | −0.622 712 | −0.012 239 |
| 6 | 1.259 777 | −2.920 928 | 0.008 209 |
| 6 | 1.259 061 | −1.537 023 | 0.012 560 |
| 1 | 2.194 471 | −3.466 983 | 0.003 709 |
| 1 | −0.004 310 | −4.680 689 | −0.005 516 |
| 6 | 1.186 839 | 2.900 893 | 0.011 068 |
| 6 | −0.054 895 | 3.536 347 | 0.004 572 |
| 1 | −3.354 417 | −1.286 269 | −0.020 883 |
| 1 | −3.374 600 | 1.173 123 | −0.020 577 |
| 1 | −2.178 473 | 3.306 425 | −0.008 537 |
| 1 | −0.101 790 | 4.617 399 | −0.001 808 |
| 1 | 2.106 260 | 3.471 148 | 0.007 660 |
| 1 | 2.698 949 | 0.004 286 | 1.079 092 |

Table 5: B3LYP/6-311++G(2d,p) optimised structure of transition state **TS5**.

| Atomic number | Coordinates (Å) | | |
| --- | --- | --- | --- |
| | X | Y | Z |
| 6 | 2.470 163 | −0.701 693 | 0.016 224 |
| 6 | 0.016 796 | 0.624 181 | −0.070 839 |
| 6 | 2.453 506 | 0.634 074 | −0.008 762 |
| 6 | −0.003 164 | −0.760 044 | −0.052 673 |
| 6 | 1.255 035 | 1.413 875 | −0.076 815 |
| 6 | 1.225 021 | −1.476 542 | −0.023 793 |
| 6 | −1.274 980 | −1.393 141 | −0.022 082 |
| 6 | 0.857 957 | 2.779 200 | −0.026 490 |
| 6 | 1.157 191 | −2.859 447 | 0.002 074 |
| 6 | −2.448 273 | −0.572 739 | 0.004 107 |
| 6 | 0.027 240 | 3.902 325 | 0.043 206 |
| 6 | −2.403 394 | 0.802 336 | 0.013 171 |
| 6 | −0.652 982 | 2.767 522 | −0.015 085 |
| 6 | −1.132 526 | 1.431 645 | −0.023 053 |
| 1 | −3.313 486 | 1.385 925 | 0.043 522 |
| 1 | −3.413 105 | −1.065 409 | 0.026 216 |
| 1 | 0.043 357 | 4.982 643 | 0.089 854 |
| 1 | 1.306 147 | 2.213 305 | −1.160 778 |
| 6 | −1.285 900 | −2.803 797 | −0.000 378 |
| 6 | −0.094 757 | −3.505 405 | 0.005 734 |
| 1 | 3.409 024 | −1.242 358 | 0.090 757 |
| 1 | 2.063 289 | −3.452 702 | 0.027 852 |
| 1 | −0.119 532 | −4.587 832 | 0.025 448 |
| 1 | −2.228 974 | −3.336 112 | 0.020 264 |

Table 6: B3LYP/6-311++G(2d,p) optimised structure of transition state **TS6**.

| Atomic number | Coordinates (Å) | | |
|:---:|:---:|:---:|:---:|
| | X | Y | Z |
| 6 | 0.027 914 | −0.590 009 | 2.447 129 |
| 6 | 0.075 890 | 0.777 121 | 2.440 231 |
| 6 | 0.076 786 | 1.456 110 | 1.184 536 |
| 6 | 0.011 819 | 0.677 466 | 0.008 989 |
| 6 | −0.028 577 | −0.707 181 | −0.011 699 |
| 6 | −0.020 033 | −1.377 928 | 1.242 292 |
| 6 | −0.051 645 | −1.401 829 | −1.265 528 |
| 6 | 0.037 532 | −0.620 569 | −2.477 352 |
| 6 | 0.041 299 | 0.743 938 | −2.446 611 |
| 6 | 0.004 584 | 1.510 593 | −1.192 305 |
| 6 | 0.120 276 | 2.789 650 | 0.738 168 |
| 6 | 0.127 464 | 3.945 105 | 0.032 124 |
| 6 | 0.078 108 | 2.842 924 | −0.751 062 |
| 6 | −0.059 176 | −2.781 744 | 1.206 201 |
| 6 | −0.093 767 | −3.460 296 | −0.007 180 |
| 6 | −0.091 506 | −2.795 320 | −1.236 348 |
| 1 | −1.030 689 | 1.296 640 | −1.981 387 |
| 1 | 0.140 826 | −1.133 673 | −3.428 259 |
| 1 | −0.121 888 | −4.542 911 | −0.001 017 |
| 1 | −0.113 315 | −3.361 786 | −2.158 913 |
| 1 | −0.063 565 | −3.342 604 | 2.133 012 |
| 1 | 0.028 321 | −1.114 984 | 3.394 503 |
| 1 | 0.111 703 | 1.333 067 | 3.367 722 |
| 1 | 0.141 384 | 5.027 535 | 0.050 031 |

Table 7: B3LYP/6-311++G(2d,p) optimised structure of transition state **TS7**.

| Atomic number | Coordinates (Å) | | |
|:---:|:---:|:---:|:---:|
| | X | Y | Z |
| 6 | 2.463 283 | 0.751 563 | 0.005 552 |
| 6 | 2.567 544 | −0.603 427 | −0.018 586 |
| 6 | 1.382 083 | −1.391 349 | −0.056 091 |
| 6 | 0.058 181 | −0.746 876 | −0.047 791 |
| 6 | 0.000 673 | 0.659 592 | −0.013 926 |
| 6 | 1.199 978 | 1.428 812 | 0.004 740 |
| 6 | −1.271 886 | 1.299 590 | −0.005 738 |
| 6 | −2.451 820 | 0.501 289 | −0.019 421 |
| 6 | −2.397 421 | −0.863 297 | −0.033 630 |
| 6 | −1.140 389 | −1.523 253 | −0.039 293 |
| 6 | 1.149 481 | −2.715 794 | −0.085 564 |
| 6 | 0.237 270 | −3.620 215 | 0.007 466 |
| 6 | −0.996 600 | −2.936 846 | 0.017 390 |
| 6 | 1.102 730 | 2.820 033 | 0.027 282 |
| 6 | −0.143 775 | 3.447 451 | 0.030 610 |
| 6 | −1.312 456 | 2.705 128 | 0.014 857 |
| 1 | −0.547 007 | −3.744 627 | −0.994 463 |
| 1 | −3.302 147 | −1.456 250 | −0.029 359 |
| 1 | −3.414 080 | 0.999 017 | −0.013 754 |
| 1 | −0.195 921 | 4.528 603 | 0.048 227 |
| 1 | −2.273 042 | 3.206 002 | 0.021 450 |
| 1 | 2.006 589 | 3.417 139 | 0.044 132 |
| 1 | 3.365 606 | 1.351 338 | 0.026 227 |
| 1 | 3.529 404 | −1.096 400 | −0.017 000 |

Table 8: B3LYP/6-311++G(2d,p) optimised structure of transition state **TS8**.

| Atomic number | Coordinates (Å) | | |
|---|---|---|---|
| | X | Y | Z |
| 6 | 0.025 596 | 0.637 905 | −2.524 888 |
| 6 | −0.005 367 | −0.693 366 | −2.553 609 |
| 6 | 0.015 542 | −1.419 718 | −1.302 602 |
| 6 | 0.031 856 | −0.691 842 | −0.068 414 |
| 6 | 0.040 384 | 0.718 062 | −0.049 715 |
| 6 | 0.088 691 | 1.438 449 | −1.352 707 |
| 6 | 0.001 189 | 1.428 087 | 1.191 783 |
| 6 | −0.010 398 | 0.687 469 | 2.409 098 |
| 6 | 0.003 230 | −0.673 782 | 2.390 690 |
| 6 | 0.016 819 | −1.406 479 | 1.168 951 |
| 6 | 0.002 165 | −2.811 884 | −1.271 466 |
| 6 | 0.001 548 | −3.496 870 | −0.057 370 |
| 6 | 0.005 801 | −2.808 461 | 1.144 922 |
| 6 | −0.014 895 | 2.873 854 | −1.194 337 |
| 6 | −0.000 817 | 3.344 082 | −0.031 354 |
| 6 | −0.020 294 | 2.869 085 | 1.227 612 |
| 1 | −0.011 160 | −3.364 578 | −2.202 924 |
| 1 | −0.008 294 | −4.579 506 | −0.056 443 |
| 1 | −0.003 790 | −3.352 932 | 2.081 285 |
| 1 | −0.042 636 | 3.421 613 | 2.157 178 |
| 1 | 1.150 306 | 2.110 965 | −1.404 576 |
| 1 | −0.058 400 | −1.246 258 | −3.485 940 |
| 1 | −0.005 052 | −1.223 667 | 3.324 291 |
| 1 | −0.030 104 | 1.225 909 | 3.348 033 |

Table 9: B3LYP/6-311++G(2d,p) optimised structure of transition state **TS9**.

| Atomic number | Coordinates (Å) | | |
|---|---|---|---|
| | X | Y | Z |
| 6 | 0.018 402 | 0.616 527 | −2.527 091 |
| 6 | 0.002 737 | −0.742 052 | −2.538 540 |
| 6 | 0.017 744 | −1.442 377 | −1.300 018 |
| 6 | 0.027 648 | −0.698 018 | −0.061 560 |
| 6 | 0.042 187 | 0.707 636 | −0.046 259 |
| 6 | 0.085 295 | 1.449 842 | −1.338 620 |
| 6 | 0.005 989 | 1.427 986 | 1.200 171 |
| 6 | −0.005 289 | 0.675 681 | 2.420 880 |
| 6 | −0.001 533 | −0.678 531 | 2.406 752 |
| 6 | 0.006 793 | −1.414 309 | 1.178 610 |
| 6 | 0.002 411 | −2.845 962 | −1.265 274 |
| 6 | −0.011 252 | −3.512 523 | −0.054 634 |
| 6 | −0.009 978 | −2.808 099 | 1.150 386 |
| 6 | 0.031 960 | 2.848 916 | −1.188 884 |
| 6 | −0.006 556 | 3.338 550 | −0.056 838 |
| 6 | −0.019 107 | 2.844 800 | 1.223 214 |
| 1 | −0.002 377 | −3.399 317 | −2.195 966 |
| 1 | −0.025 513 | −4.595 037 | −0.039 168 |
| 1 | −0.023 478 | −3.352 695 | 2.086 886 |
| 1 | −0.039 303 | 3.406 035 | 2.147 378 |
| 1 | 1.145 482 | 1.141 911 | −1.946 347 |
| 1 | −0.049 076 | −1.289 815 | −3.473 385 |
| 1 | −0.012 417 | −1.231 221 | 3.338 245 |
| 1 | −0.020 071 | 1.216 941 | 3.358 570 |

Table 10: B3LYP/6-311++G(2d,p) optimised structure of transition state **TS10**.

| Atomic number | Coordinates (Å) | | |
|---|---|---|---|
| | X | Y | Z |
| 6 | −2.476 391 | 0.927 146 | 0.033 197 |
| 6 | −2.564 451 | −0.414 097 | 0.034 699 |
| 6 | −1.498 571 | −1.294 380 | 0.016 531 |
| 6 | −0.148 032 | −0.753 316 | 0.017 407 |
| 6 | −0.016 333 | 0.659 110 | −0.000 850 |
| 6 | −1.167 455 | 1.508 245 | 0.009 054 |
| 6 | 1.283 862 | 1.247 031 | −0.014 429 |
| 6 | 2.433 795 | 0.392 542 | −0.007 102 |
| 6 | 2.325 478 | −0.957 231 | 0.016 367 |
| 6 | 1.041 686 | −1.599 066 | 0.022 616 |
| 6 | −1.407 785 | −2.697 874 | −0.070 077 |
| 6 | −0.365 570 | −3.405 089 | −0.025 264 |
| 6 | 0.974 962 | −3.006 688 | 0.022 988 |
| 6 | −0.993 599 | 2.904 010 | −0.001 695 |
| 6 | 0.273 468 | 3.454 752 | −0.019 322 |
| 6 | 1.403 228 | 2.635 201 | −0.023 552 |
| 1 | −1.108 178 | −3.454 715 | 1.009 736 |
| 1 | 1.836 726 | −3.653 187 | 0.070 851 |
| 1 | 3.210 337 | −1.580 560 | 0.024 021 |
| 1 | 3.415 340 | 0.851 790 | −0.015 080 |
| 1 | 0.392 399 | 4.530 779 | −0.028 538 |
| 1 | 2.390 618 | 3.080 933 | −0.033 707 |
| 1 | −1.866 959 | 3.544 488 | 0.004 359 |
| 1 | −3.351 064 | 1.568 259 | 0.043 751 |

Table 11: B3LYP/6-311++G(2d,p) optimised structure of transition state **TS11**.

| Atomic number | Coordinates (Å) | | |
|---|---|---|---|
| | X | Y | Z |
| 6 | −2.457 766 | 0.783 170 | −0.005 114 |
| 6 | −2.481 591 | −0.550 711 | 0.019 419 |
| 6 | −1.376 272 | −1.428 246 | 0.026 120 |
| 6 | −0.063 941 | −0.769 475 | 0.019 896 |
| 6 | 0.012 612 | 0.647 169 | −0.000 320 |
| 6 | −1.173 017 | 1.443 690 | −0.013 231 |
| 6 | 1.288 567 | 1.289 178 | 0.007 125 |
| 6 | 2.474 262 | 0.505 265 | 0.024 726 |
| 6 | 2.431 250 | −0.861 628 | 0.026 901 |
| 6 | 1.180 083 | −1.502 089 | 0.018 557 |
| 6 | −1.540 592 | −2.828 241 | 0.024 359 |
| 6 | −0.272 220 | −3.420 597 | −0.010 092 |
| 6 | 0.862 638 | −2.876 515 | −0.054 932 |
| 6 | −1.061 809 | 2.831 498 | −0.022 333 |
| 6 | 0.190 174 | 3.450 668 | −0.019 805 |
| 6 | 1.346 654 | 2.696 055 | −0.004 312 |
| 1 | 3.336 325 | −1.453 906 | 0.024 654 |
| 1 | 3.431 372 | 1.012 661 | 0.026 023 |
| 1 | 0.250 888 | 4.531 521 | −0.029 132 |
| 1 | 2.314 335 | 3.182 712 | −0.000 610 |
| 1 | −1.961 038 | 3.435 636 | −0.031 620 |
| 1 | −3.364 230 | 1.379 472 | −0.012 474 |
| 1 | −2.495 088 | −3.326 983 | 0.066 554 |
| 1 | 0.436 469 | −3.552 543 | 1.037 121 |

Table 12: B3LYP/6-311++G(2d,p) optimised structure of transition state **TS12**.

| Atomic number | Coordinates (Å) | | |
|---|---|---|---|
| | X | Y | Z |
| 6 | 0.008 960 | 2.479 103 | −0.670 830 |
| 6 | −0.052 163 | −0.011 300 | 0.723 871 |
| 6 | −0.022 260 | 2.447 570 | 0.656 631 |
| 6 | −0.035 119 | 0.000 633 | −0.686 034 |
| 6 | −0.106 868 | 1.268 238 | 1.469 082 |
| 6 | −0.013 828 | 1.232 567 | −1.412 080 |
| 6 | −0.016 408 | −1.238 477 | −1.401 154 |
| 6 | −0.002 730 | 1.135 642 | 2.888 296 |
| 6 | 0.004 099 | 1.205 895 | −2.801 151 |
| 6 | −0.005 886 | −2.470 340 | −0.662 005 |
| 6 | 0.023 815 | −0.047 892 | 3.598 313 |
| 6 | 0.008 463 | −2.401 920 | 0.676 792 |
| 6 | 0.040 174 | −1.227 522 | 2.870 424 |
| 6 | −0.004 175 | −1.231 068 | 1.456 393 |
| 1 | 0.078 944 | −2.176 437 | 3.392 353 |
| 1 | −0.002 124 | −3.414 395 | −1.196 136 |
| 1 | 0.047 148 | −0.048 395 | 4.680 985 |
| 1 | −1.185 747 | 1.291 578 | 2.080 394 |
| 6 | −0.001 406 | −1.213 756 | −2.803 758 |
| 6 | 0.004 996 | −0.010 325 | −3.487 547 |
| 1 | 0.065 198 | 3.416 003 | −1.215 933 |
| 1 | 0.021 937 | 2.137 338 | −3.353 600 |
| 1 | 0.018 459 | −0.009 728 | −4.570 253 |
| 1 | 0.011 086 | −2.149 237 | −3.349 440 |

Table 13: B3LYP/6-311++G(2d,p) optimised structure of transition state **TS13**.

Table 14: B3LYP/6-311++G(2d,p) optimised structure of transition state **TS14**.

| Atomic number | Coordinates (Å) | | |
|---|---|---|---|
| | X | Y | Z |
| 6 | −2.341 937 | 0.656 015 | −0.013 966 |
| 6 | −0.035 478 | −0.717 393 | 0.016 093 |
| 6 | −2.358 454 | −0.596 583 | −0.014 008 |
| 6 | −0.025 470 | 0.722 197 | 0.014 974 |
| 6 | −1.254 195 | −1.508 147 | 0.016 295 |
| 6 | −1.220 024 | 1.545 224 | 0.015 508 |
| 6 | 1.218 000 | 1.427 994 | −0.000 686 |
| 6 | −1.253 473 | −2.889 550 | 0.016 260 |
| 6 | −1.180 984 | 2.926 593 | 0.014 395 |
| 6 | 2.370 606 | 0.615 988 | −0.010 392 |
| 6 | −0.031 054 | −3.560 110 | 0.010 704 |
| 6 | 2.434 870 | −0.715 349 | −0.010 601 |
| 6 | 1.157 909 | −2.851 053 | 0.003 160 |
| 6 | 1.187 712 | −1.445 620 | 0.001 809 |
| 1 | 3.371 063 | −1.262 155 | −0.021 409 |
| 1 | 2.099 092 | −3.387 713 | −0.003 278 |
| 1 | −0.013 731 | −4.642 129 | 0.006 683 |
| 1 | −2.188 815 | −3.433 181 | 0.013 975 |
| 6 | 1.231 148 | 2.833 334 | 0.000 088 |
| 6 | 0.057 741 | 3.567 748 | 0.008 075 |
| 1 | −2.709 032 | 0.033 675 | 1.074 214 |
| 1 | −2.101 956 | 3.494 409 | 0.012 121 |
| 1 | 0.099 335 | 4.648 966 | 0.003 573 |
| 1 | 2.188 195 | 3.339 349 | −0.006 852 |

Table 15: B3LYP/6-311++G(2d,p) optimised structure of transition state **TS15**.

| Atomic number | Coordinates (Å) | | |
|---|---|---|---|
| | X | Y | Z |
| 6 | 0.016 945 | 0.584 948 | −2.374 832 |
| 6 | −0.042 232 | −0.702 415 | −2.537 306 |
| 6 | 0.006 405 | −1.476 138 | −1.291 543 |
| 6 | 0.043 855 | −0.719 295 | −0.089 377 |
| 6 | 0.054 856 | 0.697 139 | −0.077 563 |
| 6 | 0.155 740 | 1.552 591 | −1.365 683 |
| 6 | −0.011 300 | 1.422 465 | 1.133 171 |
| 6 | −0.023 295 | 0.693 219 | 2.372 163 |
| 6 | 0.006 950 | −0.644 721 | 2.308 244 |
| 6 | 0.031 141 | −1.436 891 | 1.154 915 |
| 6 | −0.001 326 | −2.858 997 | −1.270 325 |
| 6 | 0.010 968 | −3.538 431 | −0.044 902 |
| 6 | 0.022 926 | −2.844 268 | 1.147 072 |
| 6 | 0.005 705 | 2.964 598 | −1.265 917 |
| 6 | 0.018 623 | 3.580 331 | −0.007 308 |
| 6 | −0.043 263 | 2.840 434 | 1.147 226 |
| 1 | −0.025 767 | −3.415 583 | −2.198 802 |
| 1 | 0.001 487 | −4.620 913 | −0.039 198 |
| 1 | 0.020 749 | −3.370 655 | 2.092 616 |
| 1 | −0.052 774 | 1.239 399 | 3.307 807 |
| 1 | 0.017 410 | 4.663 532 | 0.041 140 |
| 1 | −0.105 024 | 3.340 318 | 2.107 985 |
| 1 | 1.222 904 | 2.113 592 | −1.436 098 |
| 1 | −0.117 186 | −1.197 977 | −3.500 420 |

Table 16: B3LYP/6-311++G(2d,p) optimised structure of transition state **TS16**.

| Atomic number | Coordinates (Å) | | |
| --- | --- | --- | --- |
| | X | Y | Z |
| 6 | −2.406 935 | 0.656 084 | −0.010 367 |
| 6 | −0.037 353 | −0.710 040 | −0.037 182 |
| 6 | −2.500 287 | −0.674 825 | −0.009 761 |
| 6 | 0.002 504 | 0.698 009 | −0.020 732 |
| 6 | −1.267 862 | −1.431 806 | −0.013 418 |
| 6 | −1.230 864 | 1.435 796 | −0.008 143 |
| 6 | 1.243 784 | 1.423 865 | −0.003 948 |
| 6 | −1.260 070 | −2.829 103 | 0.017 471 |
| 6 | −1.193 274 | 2.830 343 | 0.010 512 |
| 6 | 2.467 698 | 0.689 552 | 0.012 204 |
| 6 | −0.061 805 | −3.569 824 | 0.023 154 |
| 6 | 2.428 357 | −0.664 999 | 0.004 023 |
| 6 | 1.101 018 | −2.879 820 | −0.015 850 |
| 6 | 1.220 259 | −1.481 067 | −0.067 933 |
| 1 | 1.852 901 | −1.196 570 | −1.122 522 |
| 1 | 3.415 412 | 1.215 663 | 0.058 340 |
| 1 | −0.083 759 | −4.652 739 | 0.059 331 |
| 1 | −2.204 476 | −3.360 822 | 0.038 493 |
| 6 | 1.227 198 | 2.825 842 | 0.011 658 |
| 6 | 0.023 567 | 3.511 042 | 0.019 014 |
| 1 | −3.448 296 | −1.200 953 | −0.003 324 |
| 1 | −2.125 224 | 3.380 848 | 0.019 378 |
| 1 | 0.026 117 | 4.593 616 | 0.033 805 |
| 1 | 2.164 517 | 3.367 907 | 0.021 649 |

Table 17: B3LYP/6-311++G(2d,p) optimised structure of transition state **TS17**.

| Atomic number | Coordinates (Å) | | |
| --- | --- | --- | --- |
| | X | Y | Z |
| 6 | 0.003 032 | 0.627 115 | −2.505 843 |
| 6 | −0.007 174 | −0.728 627 | −2.528 744 |
| 6 | 0.005 423 | −1.448 798 | −1.298 659 |
| 6 | 0.022 127 | −0.702 386 | −0.064 570 |
| 6 | 0.039 366 | 0.702 900 | −0.044 356 |
| 6 | 0.073 529 | 1.460 314 | −1.310 027 |
| 6 | 0.013 770 | 1.426 101 | 1.182 889 |
| 6 | 0.007 333 | 0.685 048 | 2.411 455 |
| 6 | 0.007 507 | −0.670 358 | 2.415 842 |
| 6 | 0.007 611 | −1.406 545 | 1.188 182 |
| 6 | −0.011 219 | −2.852 979 | −1.276 451 |
| 6 | −0.019 953 | −3.533 434 | −0.062 638 |
| 6 | −0.012 932 | −2.779 013 | 1.084 827 |
| 6 | 0.018 050 | 2.859 799 | −1.200 882 |
| 6 | −0.019 719 | 3.556 028 | −0.040 660 |
| 6 | −0.015 533 | 2.825 101 | 1.162 527 |
| 1 | −0.021 094 | −3.402 964 | −2.209 788 |
| 1 | −0.034 791 | −4.616 645 | −0.033 427 |
| 1 | −0.038 153 | 3.364 741 | 2.102 387 |
| 1 | 1.131 206 | 1.169 924 | −1.929 133 |
| 1 | −0.050 502 | −1.263 158 | −3.471 863 |
| 1 | 0.001 015 | −1.222 208 | 3.346 595 |
| 1 | −0.000 350 | 1.235 666 | 3.343 881 |
| 1 | −0.056 137 | 4.639 263 | −0.026 324 |

Table 18: B3LYP/6-311++G(2d,p) optimised structure of transition state **TS18**.

| Atomic number | Coordinates (Å) | | |
|---|---|---|---|
| | X | Y | Z |
| 6 | 0.013 959 | 0.593 558 | −2.364 401 |
| 6 | −0.041 500 | −0.695 286 | −2.523 926 |
| 6 | 0.006 690 | −1.470 530 | −1.280 503 |
| 6 | 0.043 219 | −0.708 894 | −0.077 323 |
| 6 | 0.052 644 | 0.703 478 | −0.068 991 |
| 6 | 0.150 796 | 1.562 050 | −1.357 338 |
| 6 | −0.014 147 | 1.418 245 | 1.145 826 |
| 6 | −0.024 245 | 0.688 819 | 2.373 277 |
| 6 | 0.007 884 | −0.671 151 | 2.392 792 |
| 6 | 0.033 074 | −1.406 197 | 1.178 209 |
| 6 | −0.000 521 | −2.854 717 | −1.268 475 |
| 6 | 0.014 396 | −3.544 795 | −0.038 954 |
| 6 | 0.028 762 | −2.791 382 | 1.091 097 |
| 6 | −0.001 797 | 2.971 675 | −1.248 672 |
| 6 | 0.010 180 | 3.582 491 | 0.012 919 |
| 6 | −0.048 400 | 2.837 238 | 1.163 652 |
| 1 | −0.025 178 | −3.410 732 | −2.197 878 |
| 1 | 0.006 881 | −4.628 062 | −0.016 429 |
| 1 | −0.109 788 | 3.333 405 | 2.126 396 |
| 1 | 1.216 979 | 2.126 882 | −1.425 846 |
| 1 | −0.114 059 | −1.188 783 | −3.488 300 |
| 1 | 0.000 331 | −1.212 271 | 3.330 052 |
| 1 | −0.055 095 | 1.249 466 | 3.298 960 |
| 1 | 0.006 556 | 4.665 421 | 0.065 328 |

Table 19: B3LYP/6-311++G(2d,p) optimised structure of transition state **TS19**.

| Atomic number | Coordinates (Å) | | |
|---|---|---|---|
| | X | Y | Z |
| 6 | 2.509 572 | −0.635 072 | 0.016 572 |
| 6 | −0.017 378 | 0.672 126 | −0.002 555 |
| 6 | 2.458 559 | 0.718 432 | −0.004 508 |
| 6 | 0.024 381 | −0.743 619 | 0.014 166 |
| 6 | 1.207 752 | 1.415 662 | −0.012 775 |
| 6 | 1.313 088 | −1.429 944 | 0.020 073 |
| 6 | −1.246 586 | −1.431 689 | 0.012 277 |
| 6 | 1.071 284 | 2.782 394 | −0.018 245 |
| 6 | 1.422 801 | −2.835 715 | 0.018 188 |
| 6 | −2.475 541 | −0.750 335 | 0.023 469 |
| 6 | −0.101 615 | 3.500 576 | −0.014 374 |
| 6 | −2.477 445 | 0.617 688 | 0.024 511 |
| 6 | −1.284 774 | 2.774 134 | 0.000 194 |
| 6 | −1.268 122 | 1.363 482 | 0.007 509 |
| 1 | −2.236 649 | 3.292 381 | 0.005 949 |
| 1 | −0.110 996 | 4.583 896 | −0.021 788 |
| 6 | −0.979 601 | −2.816 299 | −0.068 066 |
| 6 | 0.139 556 | −3.392 541 | −0.020 617 |
| 1 | 3.372 086 | 1.299 521 | −0.009 958 |
| 1 | 3.463 460 | −1.146 423 | 0.025 174 |
| 1 | 2.357 440 | −3.371 501 | 0.063 298 |
| 1 | −3.419 044 | 1.153 217 | 0.028 301 |
| 1 | −3.398 891 | −1.313 457 | 0.020 937 |
| 1 | −0.575 761 | −3.510 754 | 1.022 642 |

Table 20: B3LYP/6-311++G(2d,p) optimised structure of transition state **TS20**.

| Atomic number | Coordinates (Å) | | |
|---|---|---|---|
| | X | Y | Z |
| 6 | −2.602901 | 0.487316 | 0.036326 |
| 6 | −0.011935 | −0.697047 | 0.001817 |
| 6 | −2.483399 | −0.874754 | 0.032435 |
| 6 | −0.109524 | 0.715944 | 0.023952 |
| 6 | −1.204888 | −1.493725 | 0.010919 |
| 6 | −1.440154 | 1.279468 | 0.026074 |
| 6 | 1.108233 | 1.522207 | 0.030402 |
| 6 | −1.008741 | −2.867781 | −0.002843 |
| 6 | −1.307318 | 2.683377 | −0.049794 |
| 6 | 2.370974 | 0.841034 | 0.021943 |
| 6 | 0.186286 | −3.519622 | −0.021153 |
| 6 | 2.439691 | −0.512275 | −0.004714 |
| 6 | 1.344918 | −2.726430 | −0.024199 |
| 6 | 1.265359 | −1.332833 | −0.013194 |
| 1 | 2.317787 | −3.204499 | −0.035295 |
| 1 | 0.255605 | −4.600853 | −0.031909 |
| 6 | 1.083274 | 2.932545 | 0.032550 |
| 6 | −0.247491 | 3.362348 | −0.000427 |
| 1 | −3.365093 | −1.502518 | 0.035178 |
| 1 | −3.573248 | 0.965054 | 0.037034 |
| 1 | 3.408052 | −0.998679 | −0.014274 |
| 1 | 3.274962 | 1.436109 | 0.030826 |
| 1 | 1.962329 | 3.555343 | 0.078239 |
| 1 | −0.966888 | 3.408126 | 1.044388 |

Table 21: B3LYP/6-311++G(2d,p) optimised structure of transition state **TS21**.

| Atomic number | Coordinates (Å) | | |
|---|---|---|---|
| | X | Y | Z |
| 6 | 0.008763 | 2.453628 | 0.669733 |
| 6 | −0.011467 | −0.018541 | −0.709738 |
| 6 | −0.012472 | 2.467511 | −0.684075 |
| 6 | 0.005138 | −0.014288 | 0.708147 |
| 6 | −0.016827 | 1.247121 | −1.439635 |
| 6 | 0.015809 | 1.228945 | 1.412769 |
| 6 | −0.003312 | −1.252260 | 1.425928 |
| 6 | −0.012765 | 1.314531 | −2.847425 |
| 6 | 0.021833 | 1.219950 | 2.813276 |
| 6 | −0.018536 | −2.483771 | 0.716990 |
| 6 | 0.023911 | 0.015237 | −3.365925 |
| 6 | −0.017274 | −2.518948 | −0.649169 |
| 6 | 0.067980 | −1.085071 | −2.754896 |
| 6 | −0.007814 | −1.307821 | −1.363072 |
| 1 | −0.013292 | −3.456984 | −1.187729 |
| 1 | −1.024071 | −0.700074 | −3.451275 |
| 1 | −0.020705 | −3.408643 | 1.280790 |
| 1 | −0.054019 | 2.233013 | −3.410874 |
| 6 | 0.005013 | −1.225352 | 2.841254 |
| 6 | 0.017877 | −0.003484 | 3.439885 |
| 1 | −0.020498 | 3.405710 | −1.223833 |
| 1 | 0.014804 | 3.388553 | 1.217554 |
| 1 | 0.029029 | 2.151102 | 3.368596 |
| 1 | 0.000901 | −2.151177 | 3.405187 |

Table 22: B3LYP/6-311++G(2d,p) optimised structure of transition state **TS22**.